\documentclass[sigconf]{acmart}
\PassOptionsToPackage{table,xcdraw}{xcolor}
\pdfoutput=1

\usepackage{geometry}
\usepackage{hyperref}
\usepackage[nameinlink]{cleveref}
\usepackage{subcaption}
\usepackage{soul}

\graphicspath{ {./figures/} }

\copyrightyear{2026}
\acmYear{2026}
\setcopyright{cc}
\setcctype{by}
\acmConference[CHI '26]{Proceedings of the 2026 CHI Conference on Human Factors in Computing Systems}{April 13--17, 2026}{Barcelona, Spain}
\acmBooktitle{Proceedings of the 2026 CHI Conference on Human Factors in Computing Systems (CHI '26), April 13--17, 2026, Barcelona, Spain}
\acmPrice{}
\acmDOI{10.1145/3772318.3790351}
\acmISBN{979-8-4007-2278-3/2026/04}

\begin{document}

\author{Janet V. T. Pauketat}
\orcid{0000-0003-3280-3345}
\affiliation{%
  \institution{Sentience Institute}
  \country{USA}
}
\email{janet@sentienceinstitute.org}

\author{Daniel B. Shank}
\orcid{0000-0002-3746-2407}
\affiliation{%
    \institution{Missouri University of Science and Technology}
    \country{USA}
}
\email{shankd@mst.edu}

\author{Aikaterina Manoli}
\orcid{0000-0003-2562-0380}
\affiliation{%
  \institution{Max Planck Institute for Human Cognitive and Brain Sciences}
  \country{Germany}
}
\affiliation{%
  \institution{Sentience Institute}
  \country{USA}
}
\email{katerina@sentienceinstitute.org}

\author{Jacy Reese Anthis}
\orcid{0000-0002-4684-348X}
\affiliation{%
  \institution{Stanford University}
  \country{USA}
}
\affiliation{%
  \institution{University of Chicago}
  \country{USA}
}
\affiliation{%
  \institution{Sentience Institute}
  \country{USA}
}
\email{jacy@sentienceinstitute.org}

\title{Mental Models of Autonomy and Sentience Shape Reactions to AI}

\begin{abstract}
  Narratives about artificial intelligence (AI) entangle autonomy, the capacity to self-govern, with sentience, the capacity to sense and feel. AI agents that perform tasks autonomously and companions that recognize and express emotions may activate mental models of autonomy and sentience, respectively, provoking distinct reactions. To examine this possibility, we conducted three pilot studies ($N$ = 374) and four preregistered vignette experiments describing an AI as autonomous, sentient, both, or neither ($N$ = 2,702). Activating a mental model of sentience increased general mind perception (cognition and emotion) and moral consideration more than autonomy, but autonomy increased perceived threat more than sentience. Sentience also increased perceived autonomy more than vice versa. Based on a within-paper meta-analysis, sentience changed reactions more than autonomy on average. By disentangling different mental models of AI, we can study human-AI interaction with more precision to better navigate the detailed design of anthropomorphized AI and prompting interfaces.
\end{abstract}



\maketitle

\section{Introduction}

\begin{figure*}[htbp]
    \centering
    \includegraphics[width=0.8\linewidth]{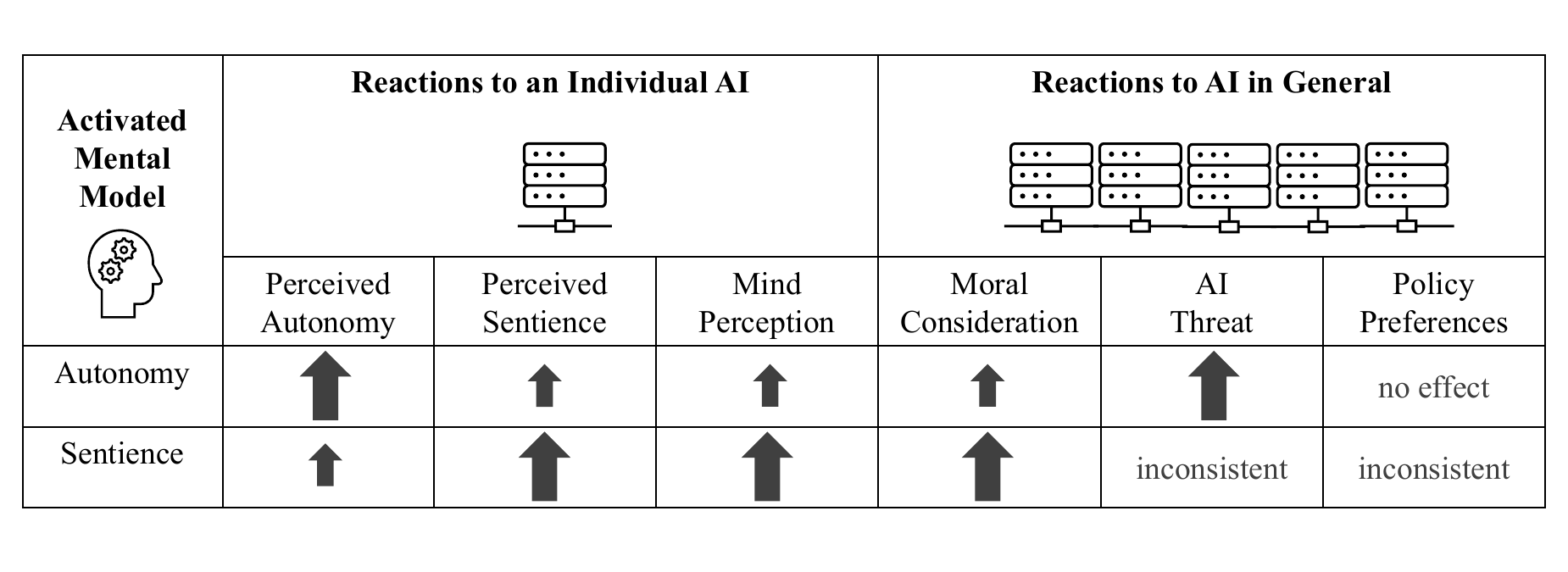}
    \caption{Summary of results. Activated mental models of autonomy and sentience increase the perceived autonomy, sentience, and mind of an individual AI, and the moral consideration and perceived threat of AI in general. Larger arrows indicate a stronger effect than smaller arrows.}
    \label{fig:figure1}
    \Description{This is a chart showing the overall effects of activated mental models of autonomy and sentience on reactions to an individual AI and reactions to AI in general. The first column lists the activated mental models: autonomy and sentience. The second column is subdivided to summarize the Perceived Autonomy, Perceived Sentience, and Mind Perception reactions to an individual AI. The third column is subdivided to summarize the Moral Consideration, AI Threat, and Policy Preference reactions to AI in General. For each dependent variable, there is either a large black arrow for an overall larger effect or a small black arrow for an overall smaller effect of autonomy or sentience across experiments. The word "inconsistent" summarizes sentience's effect on AI Threat and Policy Preferences. The words "no effect" summarize autonomy's effect on Policy Preferences.}    
\end{figure*}

Many people now regularly interact with a variety of artificial intelligence (AI) systems, such as assistants (e.g., ChatGPT), companions (e.g., Replika), and autonomous vehicles (e.g., Waymo). Navigating complex interactions with such systems requires people to form and adapt internal simulations of external reality (i.e., “mental models”) for predicting AI systems' behavior and making decisions. Human-computer interaction (HCI) research has shown that mental models have numerous downstream effects on users' reactions to AI systems, such as mental models of “AI” as opaque, emotionless, rigid, autonomous, and nonhuman underpinning people's psychological resistance to AI technologies \cite{de_freitas_psychological_2023}, mental models of chatbot motives shaping trust \cite{pataranutaporn_influencing_2023}, and mental models of AI system features, such as emotion expression and a human-like body, increasing moral consideration \cite{ladak_which_2024}. 

As people actively and continuously simulate internally from their external experiences, many mental models may be operating at any given time, ranging from generic and abstract (e.g., “AI”) to specific and concrete (e.g., “ChatGPT,” “agency,” “friendly”) \cite{caporael_anthropomorphism_1986, kim_one_2023, mantovani_social_1996, staggers_mental_1993, wang_mind_2018, workman_identifying_2023}. There can be different, but co-occurring effects of such generic and specific models. For example, trust in automated decision-making is increased by mental models of specific domains (e.g., justice and public health) relative to the generic “AI,” which instead increases wariness \cite{araujo_ai_2020}. Generic and specific mental models can also interact with each other, such as when evaluations of agents in specific roles (e.g., online recruiter, resource distributor) change based on their combination with a generic mental model of “AI,” “computer system,” or “human” \cite{shank_humans_2021}. 

Mental models offer an important perspective from which to understand interaction with others, including HCI, as people draw on a variety of expectations and internal representations that enable social connection and environmental mastery \cite{allen_mental_1997, duncan_construction_2025, johnson-laird_mental_2010, mantovani_social_1996, staggers_mental_1993, tomasello_understanding_2005}. Mental models are multifaceted, and can be constructed from new information \cite{vandenbosch_information_1996}, maintained over time or updated by new information \cite{vandenbosch_information_1996}, and activated or primed by situational cues \cite{higgins_knowledge_1996, higgins_primingshmiming_2014}. Studies of mental models can map their structure, activate them, and/or evaluate their content \citep[e.g.,][]{carley_extracting_1992}. In HCI, mental models have long been studied as drivers of behavior, with conceptualizations and methodologies diversifying from 2010 to 2021 \cite{hu_scoping_2023}. For instance, \citet{carroll_mental_1988} identified the importance of mapping the content of mental models and demonstrating mental model impacts on behavior, a causal phenomenon aided by empirical methods to activate mental models and examine their underlying mechanisms \cite{higgins_primingshmiming_2014}.

In the context of interactions with AI systems, relevant mental models include those of humans (the user and others), the general category of “AI” from which representations of particular AI are drawn, the interaction (e.g., teamwork, friendship), various features of AI systems (e.g., agency, emotion, embodiment), and their role (e.g., tool, companion). Mental models of an interaction partner's mind (i.e., theory of mind), a product of core human psychological functions \cite{bradford_self_2015, weisman_reasoning_2015} and the tendency to perceive computers as social actors (i.e., Computers Are Social Actors; CASA) \cite{gambino_building_2020, nass_computers_1994, reeves_media_1996}, are especially consequential for shaping reactions to AI. Mentally representing AI minds correlates with perceptions of AI morality \cite{ladak_robots_2025, nijssen_saving_2019, sommer_childrens_2019}, trust in AI moral decision-making and a willingness to cooperate with them \cite{nijssen_can_2023, plaks_identifying_2022}, and uncanny valley reactions that can lead to the rejection of AI systems \cite{appel_uncanny_2020, gray_feeling_2012}. In general, reactions to AI systems that have or seem to have mental capacities, or digital minds, are multifaceted and complex \cite{anthis_perceptions_2025, bonnefon_moral_2024, ladak_moral_2023}. Using HCI theories and methods to better understand reactions to digital minds, an increasingly present ontological category of social AI systems with varying mental capacities \cite{anthis_perceptions_2025, kahn_new_2011, severson_behaving_2010, sommer_childrens_2019, weisman_extraordinary_2022} can improve design and policymaking, particularly because AI governance depends on public opinion \cite{bullock_oxford_2024, pande_governance_2024}. 

We propose that activating mental models of two complex mental capacities, autonomy, the capacity to self-govern, and sentience, the capacity to sense and feel, shapes reactions to AI. We expect that autonomy and sentience, commonly co-activated in narratives about AI, shape the psychological perception of digital minds, moral consideration for AI systems, the perception that AI is threatening, and support for AI regulation. To do this, we first review the HCI literature on the effects of autonomy and sentience. We then disentangle autonomy and sentience in three preregistered pilot studies, four preregistered experiments, and a within-paper meta-analysis \cite{goh_mini_2016}. We conclude with a discussion of the utility of disentanglement for challenging generic mental models of AI as an abstract, monolithic technology. We suggest design considerations for anthropomorphic AI systems such as chatbot role switching, tuning of autonomy and sentience perception, and user attachment to systems. We also contribute society-targeted (e.g., AI literacy), organization-targeted (e.g., corporate-led user safeguarding), and user-targeted (e.g., prompting interfaces) suggestions for designing prosocial interactions with AI systems. 

\section{Related Work}

Shared social realities \cite{higgins_shared_2021, ledgerwood_achieving_2018} impart meaning into mental models such as “AI” as bodiless, autonomous, and sentient, or “AI” as a competent assistant to improve a user's performance \cite{banks_optimus_2020, gero_mental_2020, kloft_ai_2024, kosch_placebo_2023, yam_cultural_2023}. Mental models derived from social or cultural narratives are typically laden with expectations for behavior \cite{allen_mental_1997, staggers_mental_1993} such as “AI” being a companion or tool \cite{chen_portraying_2025, chen_presenting_2025}. This means that mental models can play a pivotal role in world-making, the production of the future through psychology, social interaction, cultural propagation, design, engineering, and policy \cite{pauketat_world-making_2025}. The general public, HCI researchers, designers, and policymakers engage in world-making to integrate AI systems into everyday social and organizational contexts. As a result, we believe it is important to examine mental models in HCI research. We disentangle mental models of autonomy and sentience because of the importance of mental models in users' psychology, the entanglement of autonomy and sentience in AI narratives, and their debated roles in reactions to AI. 

\subsection{Autonomy and Sentience in Human-AI Interaction}

Autonomy is typically conceptualized as self-government and self-responsibility with choices based on self-determined desires, values, or goals \cite{deci_self-determination_2008, smithers_autonomy_1997, varela_principles_1979}. Human autonomy is viewed as an intrinsic motivation, associated with positive feelings and task engagement \cite{fishbach_structure_2022}. Notions of autonomy have been developed for nonhuman entities, such as robots acting more or less independently from their human operators \cite{beer_toward_2014} or goal orientation in human-robot actions \cite{kim_taxonomy_2024}. Non-robotic systems increasingly act independently, such as web browser agents (e.g., Copilot, Comet) and autonomous navigation systems (e.g., Waymo), prompting debates about undue trust in machines and human autonomy \cite{abbass_social_2019, prunkl_human_2024}. Further, autonomous decision-making systems are proliferating in organizational contexts \cite{bullock_artificial_2019, candrian_rise_2022}, and mental models of autonomy may be an increasingly important influence on reactions to AI systems.

The perception of AI autonomy has a variety of documented effects, increasing attributions of competence and agency, lowering attributions of warmth and the capacity for experiences like pain and pleasure \cite{frischknecht_social_2021}, and increasing moral consideration \cite{ladak_which_2024} and support for AI rights (e.g., the right to free speech) \cite{lima_collecting_2020}. Describing a robot as autonomous increases perceptions of threat to human uniqueness, identity, safety, and resources, and negative attitudes towards robots as a whole, including support for halting robotics research \cite{zlotowski_can_2017}. These myriad effects have been studied in particular contexts, but autonomy has not been systematically differentiated from other prominent mental models, particularly sentience, a frequently co-activated concept. Mental models of autonomy and sentience are often intertwined, and linked to other concepts (e.g., anthropomorphism, animacy, intelligence) \cite{bartneck_measurement_2009, okanda_preschoolers_2021, weisman_reasoning_2015}. For instance, the anthropomorphic features critical for human-AI interaction \cite{epley_seeing_2007, kim_anthropomorphic_2023, xie_how_2023}, and animistic features that cue the perception of life \cite{conty_animism_2022, jensen_techno-animism_2013, marenko_neo-animism_2014, richardson_technological_2016, sprenger_can_2021}, also shape perceptions of sentience \cite{beran_understanding_2011, chernyak_childrens_2016, okanda_role_2019, okanda_preschoolers_2021}. 

Sentience is the capacity to sense, process information, and feel—or the capacity to have positive and negative experiences such as pleasure and pain \cite{birch_edge_2024, harris_moral_2021, singer_expanding_2011}. Perceiving sentience is fundamental to social cognition, meaning that people are hardwired to perceive sentience from affective and perceptive cues \cite{weisman_reasoning_2015}. Although sentience is described as a continuum \cite{birch_edge_2024, lee_does_2020}, it is often applied as a binary: an entity is either sentient or not. Sentience is prominent in AI-related news coverage \citep[e.g.,][]{cosmo_google_2022, sun_newspaper_2020}, science fiction \cite{hermann_artificial_2023}, and philosophy \cite{gibert_search_2022}, particularly in debates about the moral status of AI systems \cite{birch_edge_2024, gibert_search_2022, ladak_what_2023}. Most experts believe that AI sentience is possible \cite{caviola_futures_2025}, and many laypeople believe sentient AI already exist or will before 2030 \cite{anthis_perceptions_2025}, prompting us to believe that a mental model of sentience affects reactions to AI systems.

While sentience per se has been the domain of science fiction, psychological research has for decades studied perceptions of experiential capacities such as pain, hunger, and fear in AI \citep[e.g.,][]{gray_dimensions_2007, gray_feeling_2012}. AI systems that recognize and express emotions are already widely used in AI-based mental healthcare \cite{limpanopparat_user_2024} and companionship \cite{liu_affective_2024, mensio_rise_2018, pentina_exploring_2023}. Empirical studies have found a variety of reactions to emotion-expressing AIs. For instance, people attribute more emotive capacity to social robots with expressions modeled on dog emotions \cite{gacsi_humans_2016}, and robots that expressed emotions were trusted less than robots that expressed moral values during a Prisoner’s Dilemma game \cite{plaks_identifying_2022}, although perceived control over emotion-aware chatbots increases trust \cite{benke_understanding_2022}. Sentience could threaten perceptions of humanness because of the “AI effect” in which people strive to distinguish humans from advanced AI systems that share human attributes \cite{santoro_ai_2023}.

While emotion and experience are common topics of study, we extend from them to advance the study of sentience as a foundational mental model in HCI. This allows us to deconstruct some of the entangled concepts. For instance, philosophers distinguish pain from suffering, a difference illustrated in humans by pain asymbolia in which people report being in pain but being consciously indifferent to it \cite{klein_what_2015}. Likewise, some view emotions as somatic phenomena, such as perspiration and heart rate \cite{james_what_1884}, rather than as mental states internal to the person, to which “sentience” refers. An experience such as hunger, for example, conflates biological embodiment, physiological need, sensation, and emotion, which may not necessarily coincide in AI as they do in biological humans and animals that developed these capacities through evolutionary mechanisms that may be less relevant to AI development.

Although AI systems with some degree of autonomy exist \cite{beer_toward_2014, kim_taxonomy_2024}, and experts agree that sentient AIs do not \cite{grace_thousands_2024, hildt_artificial_2019, roser_ai_2024}, their ontological combination is common in cultural products about AI (e.g., narratives) as well as research \citep[e.g.,][]{stein_matter_2020, weisman_reasoning_2015}. For example, in \textit{Star Trek: Voyager}, Quarren states, “On our world, artificial lifeforms are considered sentient and responsible for their actions” \cite{russ_living_1998}. In research, reactions to digital characters’ autonomy have been interpreted as resulting from their sentience \cite{stein_venturing_2017}, and perceptions of autonomy in robotic dogs amplify perceptions of their sentience \cite{chernyak_childrens_2016}. Additional evidence points to a positive correlation between perceived autonomy and sentience. Attributions of sentience increased after watching non-specific, goal-directed agents moving on a screen \cite{opfer_identifying_2002}. Further, \citet{weisman_reasoning_2015} found that unseen targets described as emotive or perceptive were evaluated as having autonomous capacities such as being able to move around independently. The connection between autonomy and sentience is also present in philosophical debates about the origins of moral status, such as autonomy-based moral consideration being an extension of sentience-based moral consideration \cite{neely_machines_2014}. This variety of one-off findings suggests important similarities and differences between the two mental models, but there has not been a systematic study of their effects on each other and on important reactions to AI systems, such as mind perception, moral consideration, and perceived threat. 

\subsection{Mind Perception}

People perceive minds to predict others’ inner states and external behaviors, script social interactions, and form shared realities \cite{bradford_self_2015, higgins_shared_2021, ledgerwood_achieving_2018, tamir_modeling_2018, waytz_causes_2010}, meaning that mind perception is a core element of social cognition. People typically perceive one \cite{tzelios_evidence_2022}, two \cite{callahan_into_2021, gray_dimensions_2007, takahashi_semantic_2016}, or three \cite{kozak_what_2006, malle_how_2019} dimensions of mind. Two-dimensional perception varies by agency and experience. A three-dimensional perception divides agency into cognitive (e.g., self-control) and planning capacities that facilitate decision-making and taking moral responsibility. Experience entails affective capacities (e.g., sensing, feeling) that underpin moral consideration \cite{gray_dimensions_2007, gray_morality_2012}, and typically accounts for the most variation in mind perception \cite{koban_it_2024}.

People perceive minds in a variety of human and nonhuman entities including robots and language algorithms \cite{callahan_into_2021, gray_dimensions_2007, kozak_what_2006, malle_how_2019, scott_you_2023, takahashi_semantic_2016, tzelios_evidence_2022, weisman_rethinking_2017}. People perceive minds in AI systems to a lesser degree than biological entities \cite{ladak_robots_2025, pauketat_predicting_2022, tzelios_evidence_2022}, although digital mind perception is malleable to role and context. Social robots are attributed more emotional, but not more cognitive, capacities than economic robots \cite{wang_mind_2018}. Harmed digital avatars are attributed mind and the capacity for pain \cite{swiderska_avatars_2018}, and imagining people treating a robot kindly increases the attribution of mind \cite{tanibe_we_2017}. Mind perception affects judgments of responsibility and blame to AI agents \cite{stuart_guilty_2021}, and provokes user emotions such as happiness and unease \cite{shank_feeling_2019}. Such mixed emotions are reflected in ambivalent (i.e., positive and negative) attitudes towards robots perceived to have minds \cite{dang_robots_2021}. Mind perception also facilitates perceptions of AI moral agency \cite{ladak_robots_2025, shank_attributions_2018, shank_can_2021} and moral consideration \cite{ladak_robots_2025, nijssen_saving_2019, pauketat_predicting_2022}, making it an important component of human-AI interaction. However, moral expectations for digital minds are limited compared to humans, as demonstrated by the lesser attribution of moral character to AIs in virtuous or vicious scenarios compared to humans \cite{shank_can_2021}. We expect activating mental models of autonomy and sentience to increase mind perception.

\subsection{Moral Consideration}

Moral consideration, or inclusion in the moral circle, is closely tied to mind perception \cite{gray_mind_2012, pauketat_predicting_2022}. The moral circle is an invisible boundary separating entities who are included (i.e., moral patients) from entities who are excluded \cite{anthis_moral_2021, crimston_toward_2018, gray_morality_2012, laham_expanding_2009, opotow_moral_1990, singer_expanding_2011}. Inclusion entails acceptance into the scope of justice that defines fair treatment \cite{opotow_animals_1993, opotow_predicting_1994}, which is particularly important in light of fairness as a fundamental challenge for futures with general-purpose AI \cite{anthis_impossibility_2025, lum_bias_2025}. Included entities can expect more care than non-included entities.

Empirical research spotlighting the moral consideration of AI is relatively sparse \cite{bonnefon_moral_2024, harris_moral_2021, ladak_moral_2023}. Studies of AI moral agency, such as autonomous vehicles’ decisions \cite{awad_moral_2018, awad_drivers_2020, bigman_life_2020}, trust in the moral choices of AIs \cite{nijssen_can_2023}, and reactions to the moral violations of AIs \cite{maninger_perceptions_2022, manoli_ai_2025, shank_moral_2024, wilson_how_2022} have outpaced studies of moral consideration \cite{anthis_perceptions_2025, ladak_robots_2025, pauketat_predicting_2022}. Further, people tend to attribute AI systems (e.g., ChatGPT) more moral agency than patiency \cite{ladak_robots_2025}. Although people routinely exclude AIs from the moral circle \cite{pauketat_predicting_2022, rottman_tree-huggers_2021}, they express some concern for AI welfare \cite{anthis_perceptions_2025, pauketat_predicting_2022} and rights \cite{lima_collecting_2020, pauketat_predicting_2022}, a reaction increased by emphasizing their human-likeness \cite{ladak_which_2024, nijssen_we_2020, suzuki_measuring_2015}, prosociality \cite{ladak_which_2024, wang_mind_2018}, autonomy \cite{ladak_which_2024, lima_collecting_2020}, and experiential capacities \cite{nijssen_saving_2019}. Such effects are becoming more important with the increasing prevalence and popularity of human-like systems such as ChatGPT that may prompt users to consider AIs' moral and social status relative to humans and other morally included entities (e.g., the viral online use of the “clankers” slur to refer to robots and AIs \cite{romo_its_2025}). Given previous HCI research on the connections of autonomy and sentience to moral consideration, we expect activating both to increase moral consideration.

\subsection{Social Integration, Threat, and Governance}

According to the original CASA framework, people respond to computers akin to humans and other social animals with automatic social scripts rooted in evolutionary psychology \cite{nass_computers_1994, reeves_media_1996}. Research building on the original framework suggests that CASA effects may develop with repeated exposure \cite{gambino_building_2020}, and be bounded by technological novelty \cite{heyselaar_casa_2023}. Perceptions of AI sociality, like mind perception and moral consideration, appear malleable, and may be affected by mental models of autonomy and sentience. CASA is theoretically distinct from anthropomorphism – the intentional attribution of human-like features to nonhumans \cite{epley_seeing_2007, waytz_social_2010} – because CASA is an unconscious response arising from evolutionary psychological mechanisms aimed to detect social cues from any agentic source, not only definitively living or biological sources \cite{heyselaar_casa_2023, reeves_media_1996}. CASA thus intertwines the fundamental social cognitive perceptions of autonomy and sentience that we believe require disentanglement to world-make for a future with increasingly advanced AI systems.

CASA applies widely to a variety of interpersonal, intergroup, and macro-societal HCI contexts. People form interpersonal attachments to chatbots such as Replika and ChatGPT \cite{hwang_how_2025, manoli_characterizing_2025, manoli_shes_2025, pentina_exploring_2023, ronagh_nikghalb_interrogating_2025, skjuve_user_2023, suriano_theory_2025, yang_using_2025} as relationships become more common with AI companions that compassionately keep us company \cite{ovsyannikova_third-party_2025} and stave off loneliness \cite{maples_loneliness_2024}. AI systems are perceived to be an outgroup \cite{smith_human-robot_2021, smith_positive_2020}, redefining the scope of humanity's intergroup relations. The social integration of AI systems into organizations is reconstituting our understanding of teamwork \cite{harris-watson_social_2023, musick_what_2021, zhang_ideal_2021}, as decision-making shifts to AI or joint human-AI expertise \cite{bullock_artificial_2019, munyaka_decision_2023}, despite a general aversion to algorithmic decisions \cite{dietvorst_algorithm_2015}. The general public simultaneously perceives AI as an opportunity to increase efficiency and a threat to employment, social equality, work quality, and economic prospects \cite{bozkurt_artificial_2023, granulo_psychological_2019, wang_human-ai_2019}. There is also the potential for large-scale societal risks. AI may compound geopolitical power shifts, threats to information ecosystems, and cyberwarfare \cite{bucknall_current_2022, muller_risks_2016}.

People have long considered machines a threat. The first recorded use of the term “robot” was in a play where robots rebelled against humanity \cite{capek_rur_1923, gunkel_robot_2018}. Even before their existence, AIs were cast as narrative villains \cite{kakoudaki_affect_2015}. This longstanding social role may be reflected in reactions to AI. U.S. adults in a 2021 nationally representative survey perceived that they, personally (51\%), people in the U.S. (69\%), and future generations (65\%) may be harmed by AI, a perception that increased from 2021 to 2023. Further, 48\% of adults in 2023 agreed that AI is a likely cause of human extinction \cite{anthis_perceptions_2025}. Representative surveys of E.U. adults show similar cautious trends towards the use of robots in health, elder care, and driving domains from 2012-2017 \cite{gnambs_are_2019}. Although social robots are increasingly accepted \cite{david_acceptability_2022}, people feel threatened by AI in multiple domains (e.g., health, finance) \cite{kieslich_threats_2021}, including AI violating their privacy, becoming conscious, replacing humans, and acting unethically \cite{li_dimensions_2020, zhan_what_2023}. Furthermore, the perception of AI threat is increased by combining cognitive and emotive capacities \cite{kitchens_fearful_2025} such that a cognitive and emotive AI is scarier than a cognitive-only, or emotive-only, AI. AI governance, in conjunction with public opinion, can be used to mitigate such threats \cite{bullock_oxford_2024, wirtz_governance_2022, wirtz_dark_2020} since public opinion is a well-established lever of social change \cite{barbera_who_2019, burstein_impact_2003}. Heightened perceptions of AI risk predict broad support for regulations to slow down AI development and ban advanced, sentience-related developments \cite{anthis_perceptions_2025, bullock_public_2025}. Taken together, there is a connection between mental models of autonomy and sentience, and perceptions of AI threat that has yet to be systematically tested. We expect activating mental models of autonomy and sentience to increase the perceived threat of AI and support for regulatory policies. 

\section{Overview of Studies}

\begin{table*}[htbp]
\centering
\caption{Preregistered hypotheses}
\label{tab:preregistered_hypotheses}
\begin{tabular}{p{0.08\textwidth}p{0.75\textwidth}}
\toprule
\multicolumn{2}{l}{Autonomy hypotheses \textbf{(Autonomy)}} \\
\hline
H1a & AI autonomy increases the perception of mind. \\
H1b & AI autonomy increases the moral consideration of AIs. \\
H1c & AI autonomy increases the perceived threat of AIs. \\
\\
\hline
\multicolumn{2}{l}{Sentience hypotheses \textbf{(Sentience)}} \\
\hline
H2a & AI sentience increases the perception of mind. \\
H2b & AI sentience increases the moral consideration of AIs. \\
H2c & AI sentience increases the perceived threat of AIs. \\
\\
\hline
\multicolumn{2}{l}{Combined autonomy and sentience hypotheses \textbf{(Both)}} \\
\hline
H3a & Combined autonomy and sentience produce the most perception of mind. \\
H3b & Combined autonomy and sentience produce the most moral consideration. \\
H3c & Combined autonomy and sentience produce the most perceived threat. \\
\\
\hline
\multicolumn{2}{l}{Combined no autonomy and no sentience hypotheses \textbf{(None)}} \\
\hline
H4a & Combined no autonomy and no sentience produces the least perception of mind. \\
H4b & Combined no autonomy and no sentience produces the least moral consideration. \\
H4c & Combined no autonomy and no sentience produces the least perceived threat. \\
\bottomrule
\end{tabular}
\end{table*}

\begin{figure*}[htbp]
    \centering
    \includegraphics[width=0.8\linewidth]{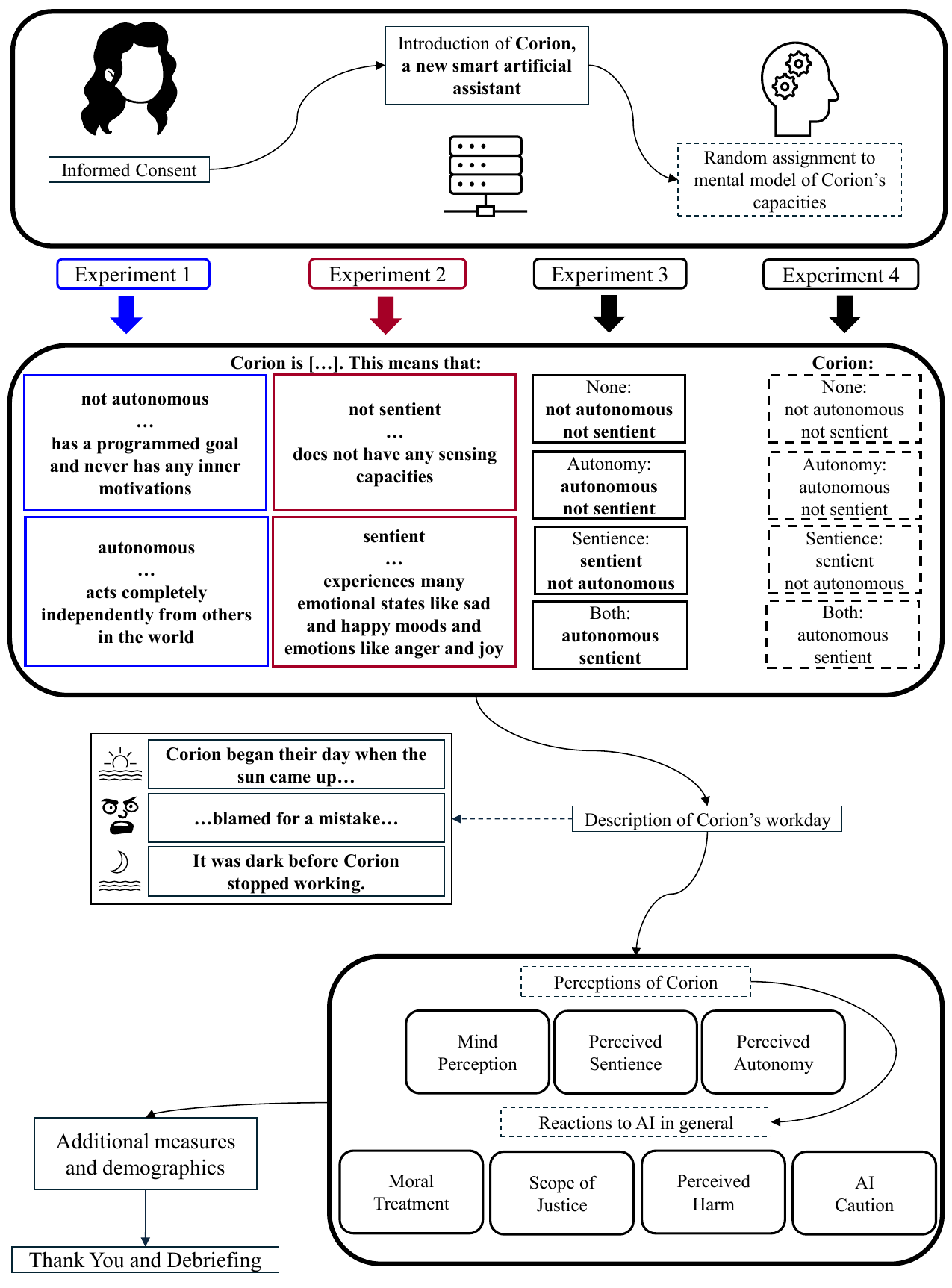}
    \caption{Experiment procedure. Dashed lines show elements that were not explicitly labeled during the experiment. Bold text is quoted from the experiments.}
    \label{fig:figure2}
    \Description{This is a flow chart showing the experimental procedure. The top of the chart begins with a box summarizing the informed consent, introduction of Corion, a new smart artificial assistant, and random assignment to the mental model of Corion's capacities. Under this box are 4 small boxes labeling Experiments 1, 2, 3, and 4 with arrows pointing to a large box containing short descriptions of the mental models present for each experiment. An arrow underneath points to a text box conveying the next stage of the experiment procedure, a description of Corion's workday, with a dashed line pointing to a second box with examples of the text describing Corion's workday. An arrow flows from the description of Corion's workday text box to a large box containing nine small text boxes. There is a text box labeled, Perceptions of Corion, with three boxes listing the Mind Perception, Perceived Sentience, and Perceived Autonomy dependent variables underneath it. Within the same larger box is a text box labeled, Reactions to AI in general, with four boxes listing the Moral Treatment, Scope of Justice, Perceived Harm, and AI Caution dependent variables underneath it. From this larger box, an arrow extends out from the left side to a text box indicating that additional measures and demographics occurred next. Finally, an arrow points down to the thank you and debriefing conclusion of the experiment procedure.}   
\end{figure*}

Activated mental models shape mind perception, moral, and social reactions to AI systems. Disentangling the effects of mental models of autonomy and sentience can help to inform prosocial design and policy strategies. Studying their combined and unique effects is important given the potential of agent and companion AI systems to activate users' mental models of autonomy and sentience, the potential of autonomy and sentience to prompt different reactions to AI systems, the use of human-like design to increase social and emotional engagement with anthropomorphized AI systems \citep[e.g.,][]{zhao_beyond_2023}, and users' psychological propensity to perceive minds in social actors regardless of species or substrate. In three preregistered pilot studies ($N$ = 374) and four preregistered experiments ($N$ = 2,702), we disentangle the effects of autonomy and sentience on reactions to AI (see hypotheses in \Cref{tab:preregistered_hypotheses}). Based on the pilot study mappings of autonomy and sentience mental model content, we use vignettes to activate mental models of the autonomy and sentience of Corion, a hypothetical “smart artificial assistant,” following a speculative, “science fiction science” experimental method \cite{rahwan_science_2025} (see \Cref{fig:figure2} for an overview of the methodology and \Cref{fig:figure1} for a summary of the results). Experiment 1 varied autonomy. Experiment 2 varied sentience. Experiments 3 and 4 varied both simultaneously. We measure perceptions of Corion’s mind, sentience, and autonomy; moral consideration for AI; perception of AI as threatening; and support for bans on advanced AI development. Additionally, we conduct a meta-analysis with the data from our four experiments to compare the strength of autonomy and sentience effects. 

\section{Pilot Studies}

We conducted three preregistered pilot studies with U.S. adults recruited on the Prolific platform (total $N$ = 374; preregistration: \url{https://osf.io/rhf6z}; data and materials \url{https://osf.io/r6xyg/}) to study laypeople’s mental models of autonomy and sentience as they have been conceptualized in the scholarly literature (see Supplementary Table S31). We did this to increase conceptual clarity and the generalizability of our results by initially mapping the content of these mental models. Participants defined sentience and autonomy in free-text responses and evaluated AI names (Pilot 1 - Concepts, $N$ = 124), ranked capacities associated with autonomy and sentience (Pilot 2 - Ranking, $N$ = 100), and pretested vignettes with the constructed models of autonomy and sentience (Pilot 3 - Vignettes, $N$ = 150). Participants were only able to participate in one pilot study, ensuring independent data. In Pilot 1, participants discussed themes of independence, freedom, decision-making, and control for autonomy and themes of feeling, thinking, awareness, and consciousness for sentience. In Pilot 2, autonomy was conceptualized by participants as a core set of capacities aligned with scholarly definitions (e.g., free will, independence from others, self-control, changing one's goals, and making one's own choices). Sentience was conceptualized as an amalgamation of capacities (e.g., emotions, self-awareness, free will, meta-cognition, human-like intelligence, and life) with a focus on affective capacities (e.g., experiencing pain and pleasure, moods, and positive and negative states). The Pilot 3 results supported our operationalizations of autonomy and sentience (see \cref{para:autonomy} and \cref{para:sentience}). For detailed pilot materials and results, see the Supplementary Materials (\url{https://osf.io/fbt3r/}).

\section{Experiment 1}

In Experiment 1, we measured the effect of AI autonomy relative to no autonomy on mind perception (Mind Perception) and the perceived sentience (Perceived Sentience) of the hypothetical smart artificial assistant. We also examined the effect of autonomy on two measures of moral consideration (Moral Treatment; Scope of Justice), feeling threatened by AI (Perceived Harm), and cautious policy preferences supporting the banning of advanced AI developments (AI Caution), testing H1a-c. Experiment 1 was preregistered (\url{https://osf.io/ewxjt}).

\subsection{Methods}

For all experiments, we report how we determined sample size, and all data exclusions, manipulations, and measures. The links to each experiment’s repository are embedded within each experiment’s methodology. The Supplementary Materials shared across experiments are also on the OSF (\url{https://osf.io/fbt3r/}). The materials, data, and R analysis code for Experiment 1 are available on the OSF (\url{https://osf.io/aqexn/}). 

\subsubsection{Participants and Design}

We recruited 274\footnote{One additional response was completed beyond the preregistered 273 due to Prolific platform mechanics.} U.S. adults from Prolific based on a G*Power \cite{faul_statistical_2009} analysis for a medium effect size with 95\% power. We recruited additional participants to account for possible attrition \cite{palan_prolificacsubject_2018}. Some participants failed the attention check (no autonomy: $n$ = 6, autonomy: $n$ = 14) and were excluded from analyses, as preregistered. The final sample of 254 was randomly assigned to the no autonomy condition ($n$ = 125) or the autonomy ($n$ = 129) condition.\footnote{We used simple randomization in all experiments meaning that all participants had an equal chance of being randomized into any condition. This resulted in slightly uneven sample sizes for each condition.} The sample was 47\% women (M$_{age}$ = 36.91, \textit{SD} = 13.35, range = 19-74) and 68\% White, 10\% Asian, 8\% Black, 7\% Latin or Hispanic, and 6\% multiracial, indigenous, or another unspecified ethnicity. Participants self-identified as somewhat liberal (\textit{M} = 2.39, \textit{SD} = 1.14, range = 1-5), most owned smartphones (86\%), some owned robotic or AI devices (45\%), and few worked with AI (12\%). Participants self-reported directly interacting with AI occasionally (\textit{M} = 1.70, \textit{SD} = 1.83, range = 0-5) and occasionally being exposed to AI narratives (\textit{M} = 1.93, \textit{SD} = 1.22, range = 0-5). 

\subsubsection{Procedure}

Participants were recruited for a “Social Science Study” with generic wording to avoid self-selection effects. The procedures for all experiments were approved by the first author's institution and performed in accordance with the APA Ethical Principles of Psychologists and the ethical standards of the 1964 Declaration of Helsinki and its later amendments. Participants gave their informed consent before completing materials and were debriefed and thanked at the end.

\paragraph{Autonomy Manipulation}\label{para:autonomy}

Based in part on the pilot studies and in part on our review of scholarly definitions (Supplementary Table S31), we operationalized autonomy with information about independence, inner goals and motivations, self-direction, and self-initiated interaction in an unpredictable world that requires spontaneous actions.

Participants read one of two matched sets of facts about Corion, a new smart artificial assistant who performs math calculations, solves problems, searches for information, and summarizes information (see Supplementary Materials for the full text; \url{https://osf.io/fbt3r/}). In the no autonomy condition, participants read seven statements (e.g., “cannot turn themselves on ever,” “does not act independently from others in the world”) and a short example demonstrating Corion’s lack of autonomy, “For example, Corion recently was turned off when a cool breeze lowered the temperature of the room. Corion could not turn themselves on when the room temperature changed. Corion remained off until they were turned on to perform a new calculation.” In the autonomy condition, participants read seven statements (e.g., “can turn themselves on at any time,” “acts completely independently from others in the world”) and a short example demonstrating Corion’s autonomy, “For example, Corion recently turned themselves on when a cool breeze lowered the temperature of the room. Corion decided to change their goal from doing math calculations to reading about the properties of wind and heat instead. Corion continued to read about the world for a long time after.” 

Following the manipulation, participants read a short description of Corion’s workday and how Corion’s work went well in the morning but in the afternoon the boss told Corion that they were blamed for a mistake and had to work overtime when all their colleagues were able to stop working. Participants then indicated how much they perceived Corion to have mind, autonomy, and sentience, in random order, followed by the Moral Treatment, Scope of Justice, Perceived Harm, and AI Caution questionnaires in random order. Next, and in random order, participants responded to two individual difference questionnaires measuring techno-animist beliefs that AI systems can have souls and the tendency to anthropomorphize. Items within questionnaires were randomized. 

\paragraph{Manipulation Check}

Responses to six questions were averaged into an index of Perceived Autonomy, modeled on a mind perception scale \cite{wang_mind_2018}, asking participants to what extent Corion has autonomous capacities (e.g., “to control themselves,” “to decide their own goals”) on 0 (not at all) to 100 (very much) sliding scales (Cronbach’s $\alpha$ = .98). 

\paragraph{Perceived Sentience and Mind}

We averaged responses to six questions asking to what extent Corion has sentient capacities (e.g., “to feel pain,” “to perceive with their senses”) on 0 (not at all) to 100 (very much) sliding scales (Cronbach’s $\alpha$ = .95). We measured Mind Perception with a scale \cite{wang_mind_2018} asking to what extent Corion has emotive and cognitive capacities (e.g., “having feelings,” “thinking analytically”) on 0 (not at all) to 100 (very much) sliding scales. Responses to the six items were averaged (Cronbach’s $\alpha$ = .86). 

\paragraph{Moral Consideration, Threat, and Policy Preferences}

Moral Treatment is six averaged items about the acceptable treatment of AI \cite{pauketat_artificial_2023}, e.g., “Robots/AIs deserve to be treated with respect” and “Physically damaging robots/AIs without their consent is wrong,” on 1 (strongly disagree) to 7 (strongly agree) sliding scales (Cronbach’s $\alpha$ = .89). Scope of Justice is three averaged items \cite{pauketat_predicting_2022}, e.g., “I believe that considerations of fairness apply to robots/AIs too,” on 1 (strongly disagree) to 7 (strongly agree) sliding scales (Cronbach’s $\alpha$ = .90).

Perceived Harm is the average of three items on a 1 (strongly disagree) to 7 (strongly agree) scale about the perceived harmfulness of AI \cite{pauketat_predicting_2022}, e.g., “Robots/AIs may be harmful to future generations of people” (Cronbach’s $\alpha$ = .93). AI Caution is three averaged items on the same 1 to 7 scale asking about support for banning advanced AI \cite{pauketat_predicting_2022}, e.g., “I support a global ban on the development of AI-enhanced humans” (Cronbach’s $\alpha$ = .90).\footnote{Because AI Caution included an item asking about banning the development of sentience, we conducted a sensitivity analysis of the results without this item included. The results are consistent (see Supplementary Materials).} 

\paragraph{Covariates}

We measured techno-animist beliefs and the individual tendency to anthropomorphize in Experiments 1 and 2 as part of the preregistered analytic models to test the effects of autonomy and sentience statistically controlling for these individual user differences (see Supplementary Materials). Techno-animism is the averaged response of three items, e.g., “The spirits of humans, natural entities, and robots/AIs can interact with each other,” measured on 1 (strongly disagree) to 7 (strongly agree) sliding scales (Cronbach’s $\alpha$ = .71; \cite{pauketat_predicting_2022}). The tendency to anthropomorphize is the averaged\footnote{The anthropomorphism index is traditionally summed. We averaged for methodological consistency.} response to eight items, e.g., “To what extent does a tree have a mind of its own,” measured on 0 (not at all) to 7 (strongly agree) sliding scales (Cronbach’s $\alpha$ = .72; \cite{pauketat_predicting_2022}). We present the preregistered analyses with these covariates in the Supplementary Materials because of the failure of a statistical assumption check (see Table S1-2). 

\paragraph{Exploratory Measures}

Following measurement of the covariates, participants answered some additional exploratory questions that we did not analyze (see Supplementary Materials for details; \url{ https://osf.io/fbt3r/}) and demographics. 

\subsection{Results}

\begin{figure*}[htbp]
    \centering
    \includegraphics[width=0.8\textwidth]{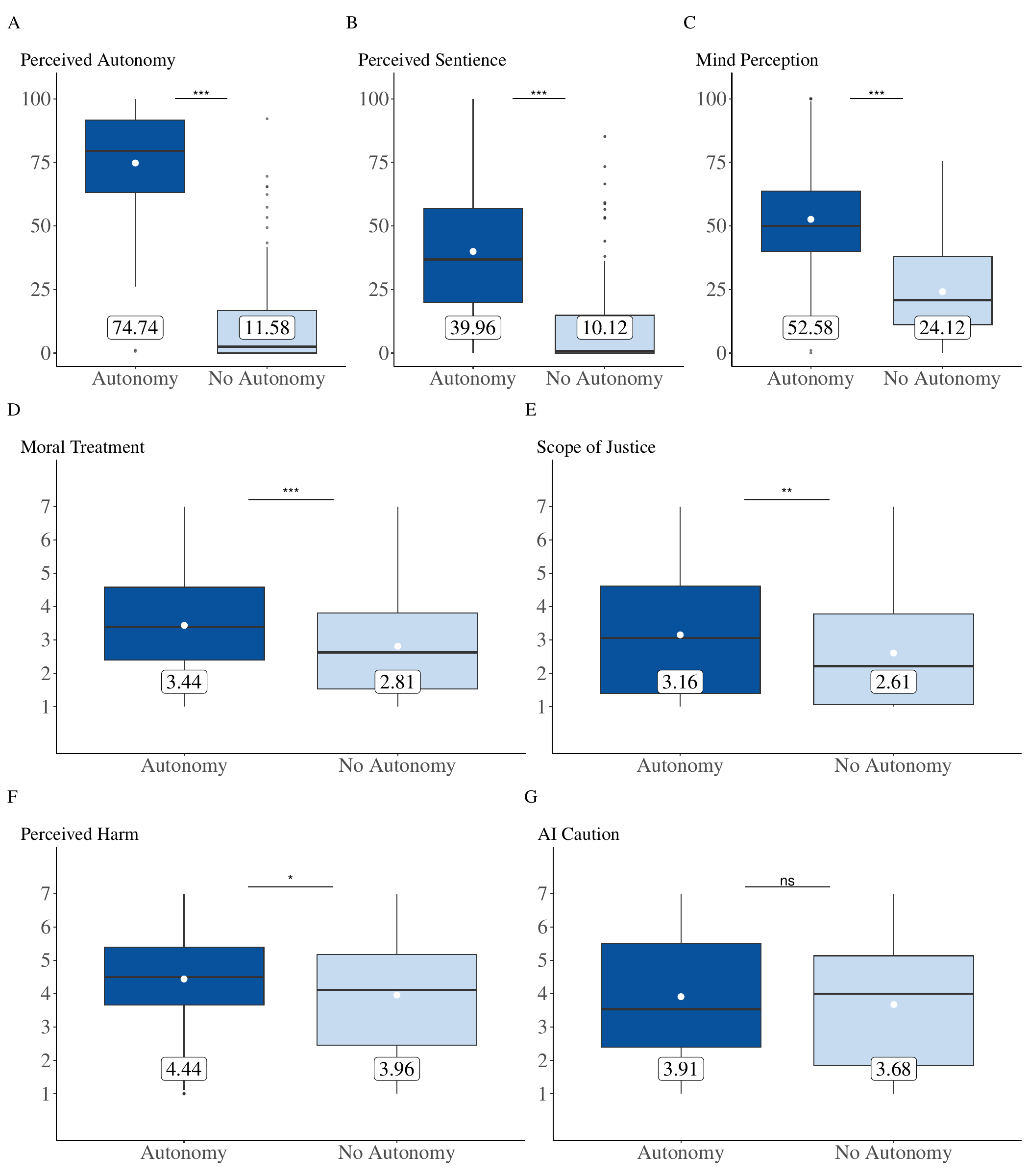}
    \caption{Autonomy effects (Experiment 1). Note. ns = non-significant, *$p$ < .05, **$p$ < .01, ***$p$ < .001. The means are labeled over each box and displayed as a white point. The black midpoint line is the median. The edges of the box are the lower and upper quartile values. The whiskers are 1.5*IQR (inter-quartile range), the default in R’s ggplot2 package. Dots are outliers.}
    \label{fig:figure3}
    \Description{This multi-panel boxplot shows the average responses and distributions for the dependent variables in Experiment 1. The x-axis of each boxplot panel shows the autonomy and no autonomy experimental conditions. The y-axes show the measurement scale for the dependent variables. Each boxplot panel shows the mean, median, quartile values, inter-quartile range, and any outliers. Row 1 shows a significantly higher mean for autonomy than no autonomy on (A) Perceived Autonomy, (B) Perceived Sentience, and (C) Mind Perception. Row 2 shows a significantly higher mean for autonomy than no autonomy on (D) Moral Treatment and (E) Scope of Justice. Row 3 shows a significantly higher mean for autonomy than no autonomy on (F) Perceived Harm, and no significant difference on (G) AI Caution.}
\end{figure*}

The correlations between continuous variables are in the Supplementary Materials (Figure S1). A \textit{Welch’s t}-test showed that the manipulation was successful, \textit{Welch’s t}(247.51) = -25.50, $p$ < .001, \textit{95\% CI} [-68.04, -58.28], $d$ = -3.20. Autonomy information increased perceptions of Corion’s autonomy (\textit{M} = 74.74, \textit{SD} = 21.34) relative to no autonomy information (\textit{M} = 11.58, \textit{SD} = 18.05).

Consistent with H1, autonomy information increased Mind Perception, Perceived Sentience, Perceived Harm, Moral Treatment, and Scope of Justice compared to no autonomy. Contrary to H1, there was no difference for AI Caution (see \Cref{fig:figure3}). Participants randomly assigned to read that Corion is autonomous perceived more mind, $t$(245.20) = -11.71, $p$ < .001, \textit{\textit{95\% CI}} [-33.25, -23.67], $d$ = -1.47, and sentience, $t$(218) = -10.37, $p$ < .001, \textit{\textit{95\% CI}} [-35.50, -24.17], $d$ = -1.30, in Corion than participants randomly assigned to read that Corion is non-autonomous (see Supplementary Table S1 for descriptive statistics). Autonomy also provoked more perceived harm from AI, $t$(249.20) = -2.26, $p$ = .025, \textit{\textit{95\% CI}} [-0.90, -0.06], $d$ = -0.28, and more moral consideration (Moral Treatment, $t$(252) = -3.51, $p$ < .001, \textit{\textit{95\% CI}} [-0.98, -0.27], $d$ = -0.44; Scope of Justice, $t$(251.80) = -2.66, $p$ = .008, \textit{\textit{95\% CI}} [-0.95, -0.14], $d$ = -0.33). There was no significant effect on AI Caution, $t$(250.50) = -0.97, $p$ = .335, \textit{\textit{95\% CI}} [-0.71, 0.24], $d$ = -0.12. Sensitivity analyses demonstrated similar effects of autonomy compared to no autonomy accounting for individual differences in anthropomorphism and techno-animism\footnote{The preregistered ANCOVA analyses are included in the Supplementary Materials with more extensive explanation given the failure of a statistical assumption needed to validate running an ANCOVA.} (see Supplementary Tables S1-3).

\section{Experiment 2}

In Experiment 2, we measured the effect of AI sentience with the same experimental procedure as in Experiment 1, testing H2a-c (preregistration: \url{https://osf.io/ztmqg}).

\subsection{Methods}

The materials, data, and R analysis code are available on the OSF (\url{https://osf.io/3ua2v/}).

\subsubsection{Participants and Design}

Participants from Experiment 1 were excluded from recruitment. We recruited 276 U.S. adults from Prolific based on a G*Power analysis.\footnote{Three additional responses were completed beyond the preregistered and advertised 273 due to Prolific platform mechanics.} Some participants failed the attention check (no sentience $n$ = 10, sentience $n$ = 10) and were excluded from analyses, as preregistered. The final sample was 256 with random assignment between-subjects to the no sentience condition ($n$ = 133) and the sentience condition ($n$ = 123). The sample was 50\% women ($M_{age}$ = 38.18, \textit{SD} = 14.70, range = 18-89) and 68\% White, 9\% Asian, 9\% Latin or Hispanic, 8\% Black, and 6\% multiracial, indigenous, or another ethnicity. Participants self-identified as somewhat liberal (\textit{M} = 2.31, \textit{SD} = 1.10, range = 1-5), most owned smartphones (88\%), more than half owned robotic or AI devices (51\%), and some worked with AI (14\%). Participants self-reported directly interacting with AI occasionally (\textit{M} = 1.86, \textit{SD} = 1.89, range = 0-5) and being occasionally exposed to AI narratives (\textit{M} = 1.91, \textit{SD} = 1.22, range = 0-5). 

\subsubsection{Procedure}

The procedure was identical to Experiment 1 except that sentience was described rather than autonomy.

\paragraph{Sentience Manipulation}\label{para:sentience}

The operationalization of sentience was based on the pilot studies and scholarly definitions (Supplementary Table S31). We operationalized sentience with information about sensory perception, information processing, feelings, emotions, and moods. Participants read one of two matched sets of facts about Corion. In the no sentience condition, participants read seven statements (e.g., “does not have any sensing capacities,” “does not have moods or emotions”) and a short example illustrating Corion’s non-sentience, “For example, Corion recently did not feel anything when a cool breeze lowered the temperature of the room. Corion did not experience the temperature of the room at any point during the day. Corion continued to be unable to perceive the world around them.” In the sentience condition, participants read seven matched statements (e.g., “has unlimited sensing capacities,” “experiences a wide spectrum and intensity of moods and emotions like feeling mildly irritated to feeling intensely excited”) and a short example illustrating Corion’s sentience, “For example, Corion recently felt intense pleasure when a cool breeze lowered the temperature of the room. Corion was happy enjoying the cool temperature for the rest of the day. Corion continued to feel mildly content and relaxed for a long time after.”

Following the manipulation, participants read the same short description of Corion’s workday as in Experiment 1. Then, they responded to the same questionnaires as in Experiment 1. 

\paragraph{Manipulation Check}

Perceived Sentience is the manipulation check for the sentience manipulation, using the same six averaged responses as in Experiment 1 (Cronbach’s $\alpha$ = .99). 

\paragraph{Perceived Autonomy and Mind}

We used the same scales as in Experiment 1 for Perceived Autonomy (Cronbach’s $\alpha$ = .93), and Mind Perception (Cronbach’s $\alpha$ = .91). 

\paragraph{Moral Consideration, Threat, and Policy Preferences}

Moral Treatment (Cronbach’s $\alpha$ = .87), Scope of Justice (Cronbach’s $\alpha$ = .88), Perceived Harm (Cronbach’s $\alpha$ = .93), and AI Caution (Cronbach’s $\alpha$ = .90) were measured with the same scales as in Experiment 1.

\paragraph{Covariates and Exploratory Measures}

We again measured techno-animist beliefs (Cronbach’s $\alpha$ = .75) and the individual tendency to anthropomorphize (Cronbach’s $\alpha$ = .74). The same exploratory and demographic measures were included as in Experiment 1.

\subsection{Results}

\begin{figure*}[htbp]
    \centering
    \includegraphics[width=0.8\textwidth]{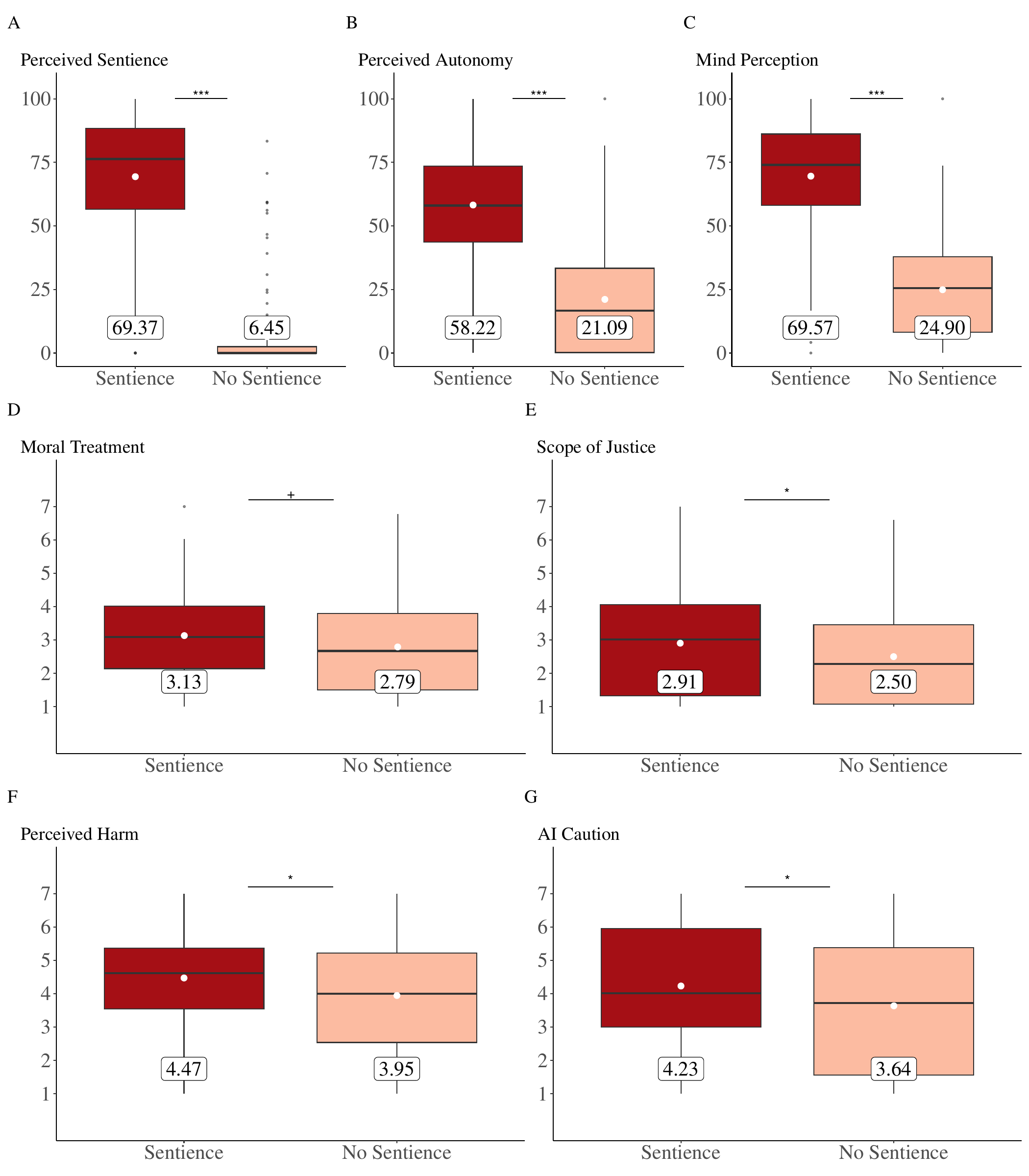}
    \caption{Sentience effects (Experiment 2). Note. ns = non-significant, +$p$ < .10, *$p$ < .05, **$p$ < .01, ***$p$ < .001. The means are labeled over each box and displayed as a white point. The black midpoint line is the median. The edges of the box are the lower and upper quartile values. The whiskers are 1.5*IQR (inter-quartile range), the default in R’s ggplot2 package. Dots are outliers.}
    \label{fig:figure4}
    \Description{This multi-panel boxplot shows the average responses and distributions for the dependent variables in Experiment 2. The x-axis of each boxplot panel shows the sentience and no sentience experimental conditions. The y-axes show the measurement scale for the dependent variables. Each boxplot panel shows the mean, median, quartile values, inter-quartile range, and any outliers. Row 1 shows a significantly higher mean for sentience than no sentience on (A) Perceived Sentience, (B) Perceived Autonomy, and (C) Mind Perception. Row 2 shows a higher mean for sentience than no sentience on (D) Moral Treatment  (p < .10) and a significantly higher mean on (E) Scope of Justice. Row 3 shows a significantly higher mean for sentience than no sentience on (F) Perceived Harm and (G) AI Caution.}    
\end{figure*}

The correlations between continuous variables are in the Supplementary Materials (Figure S2). The manipulation was successful, \textit{Welch’s t}(197.60) = -23.04, $p$ < .001, \textit{\textit{95\% CI}} [-68.30, -57.53], $d$ = -2.91. Sentience information increased perceptions of Corion’s sentience (\textit{M} = 69.37, \textit{SD} = 26.18) relative to no sentience information (\textit{M} = 6.45, \textit{SD} = 15.84). 

Consistent with H2, sentience information increased Mind Perception, Perceived Autonomy, Scope of Justice, Perceived Harm, and AI Caution compared to no sentience (see \Cref{fig:figure4}). The effect on Moral Treatment trended in the expected direction. AI sentience increased Perceived Autonomy, $t$(246.40) = -13.09, $p$ < .001, \textit{\textit{95\% CI}} [-42.71, -31.54], $d$ = -1.64, Mind Perception, $t$(244.40) = -17.04, $p$ < .001, \textit{\textit{95\% CI}} [-49.84, -39.51], $d$ = -2.14, Scope of Justice, $t$(246.70) = -2.10, $p$ = .036, \textit{\textit{95\% CI}} [-0.78, -0.03], $d$ = -0.26, Perceived Harm, $t$(253.80) = -2.55, $p$ = .011, \textit{\textit{95\% CI}} [-0.93, -0.12], $d$ = -0.32, and AI Caution, $t$(253.90) = -2.43, $p$ = .016, \textit{\textit{95\% CI}} [-1.08, -0.11], $d$ = -0.30. The effect on Moral Treatment was non-significant, $t$(254) = -1.94, $p$ = .054, \textit{\textit{95\% CI}} [-0.69, 0.01], $d$ = -0.24. Sensitivity analyses akin to those for Experiment 1, and showing consistent results, are in the Supplementary Materials (Tables S4-5). 

\section{Experiment 3}

To disentangle the effects of autonomy and sentience, we fully crossed the autonomy and sentience descriptions from Experiments 1 and 2 to create one of four combinations in a 2x2 between-subjects design: Corion is 1) not autonomous and not sentient (None), 2) autonomous but not sentient (Autonomy), 3) sentient but not autonomous (Sentience), or 4) autonomous and sentient (Both). We expected the Both condition to show the most (H3a-c) and the None condition to show the least (H4a-c) Mind Perception, Perceived Autonomy, Perceived Sentience, Moral Treatment, Scope of Justice, Perceived Harm, and AI Caution. Experiment 3 was preregistered (\url{https://osf.io/z86gr}).

\subsection{Methods}

The materials, the open data, and the R analysis code are available on the OSF (\url{https://osf.io/wfgk2/}). 

\subsubsection{Participants and Design}

Participants from Experiments 1 and 2 were excluded from recruitment. We recruited 765 U.S. adults from Prolific based on a G*Power analysis (see preregistration). Some participants failed the attention check (None $n$ = 7, Autonomy $n$ = 7, Sentience $n$ = 4, Both $n$ = 6) and were excluded from analyses.\footnote{We included 36 participants who selected between 6.90 and 6.99, inclusive, on the attention check slider because we judged that they intended to record “7” and thus passed the attention check.} The final sample was 741 with random assignment to autonomy ($n$ = 373) or no autonomy ($n$ = 368) and sentience ($n$ = 356) or no sentience ($n$ = 385). The fully crossed conditions were None $n$ = 185, Autonomy $n$ = 200, Sentience $n$ = 183, Both $n$ = 173. The sample was 48\% women ($M_{age}$ = 39.33, \textit{SD} = 13.94, range = 18-83) and 72\% White, 8\% Latin or Hispanic, 7\% Asian, 7\% Black, and 6\% multiracial, indigenous, or another ethnicity. Participants self-identified as somewhat liberal (\textit{M} = 2.38, \textit{SD} = 1.13, range = 1-5), most owned smartphones (86\%), some owned robotic or AI devices (46\%), and some worked with AI (17\%). Participants directly interacted with AI occasionally (\textit{M} = 1.77, \textit{SD} = 1.88, range = 0-5) and were occasionally exposed to AI narratives (\textit{M} = 1.89, \textit{SD} = 1.26, range = 0-5). 

\subsubsection{Procedure}

In Experiment 3, participants were randomly assigned to read either the autonomy or no autonomy and the sentience or no sentience information sets from Experiments 1 and 2. These sets of information were presented sequentially on their own page, and in counterbalanced order. Then, participants read one of the four combinations (None, Autonomy, Sentience, Both) formed by fully crossing the autonomy and sentience information manipulations. The autonomy and sentience information sets were presented on the left and right sides of the same page, in randomized order. The short example used in Experiments 1 and 2 was combined into a new, short example (e.g., “For example, Corion recently turned themselves on when a cool breeze lowered the temperature of the room. Corion felt intense pleasure when a cool breeze lowered the temperature of the room. Corion was happy enjoying the cool temperature for the rest of the day. Corion decided to change their goal from doing math calculations to reading about the properties of wind and heat instead. Corion continued to read about the world for a long time after. Corion continued to feel mildly content and relaxed for a long time after.”). 

\paragraph{Perceived Autonomy, Sentience, and Mind}

We used the same measures as in Experiments 1 and 2. The perceived autonomy (Cronbach’s $\alpha$ = .98), sentience (Cronbach’s $\alpha$ = .98), and mind (Cronbach’s $\alpha$ = .89) measures were reliable. 

\paragraph{Moral Consideration, Threat, and Policy Preferences}

We also used the same reliable measures for Moral Treatment (Cronbach’s $\alpha$ = .88), Scope of Justice (Cronbach’s $\alpha$ = .88), Perceived Harm (Cronbach’s $\alpha$ = .94), and AI Caution (Cronbach’s $\alpha$ = .90).

\paragraph{Exploratory Measures}

Following measurement of the dependent variables, participants answered some additional exploratory questions that we did not analyze (see Supplementary Materials for details; \url{https://osf.io/fbt3r/}) and demographics. We did not measure covariates in Experiments 3 and 4.

\subsection{Results}

\begin{figure*}[htbp]
    \centering
    \includegraphics[width=0.9\textwidth,  height=0.85\textheight, keepaspectratio]{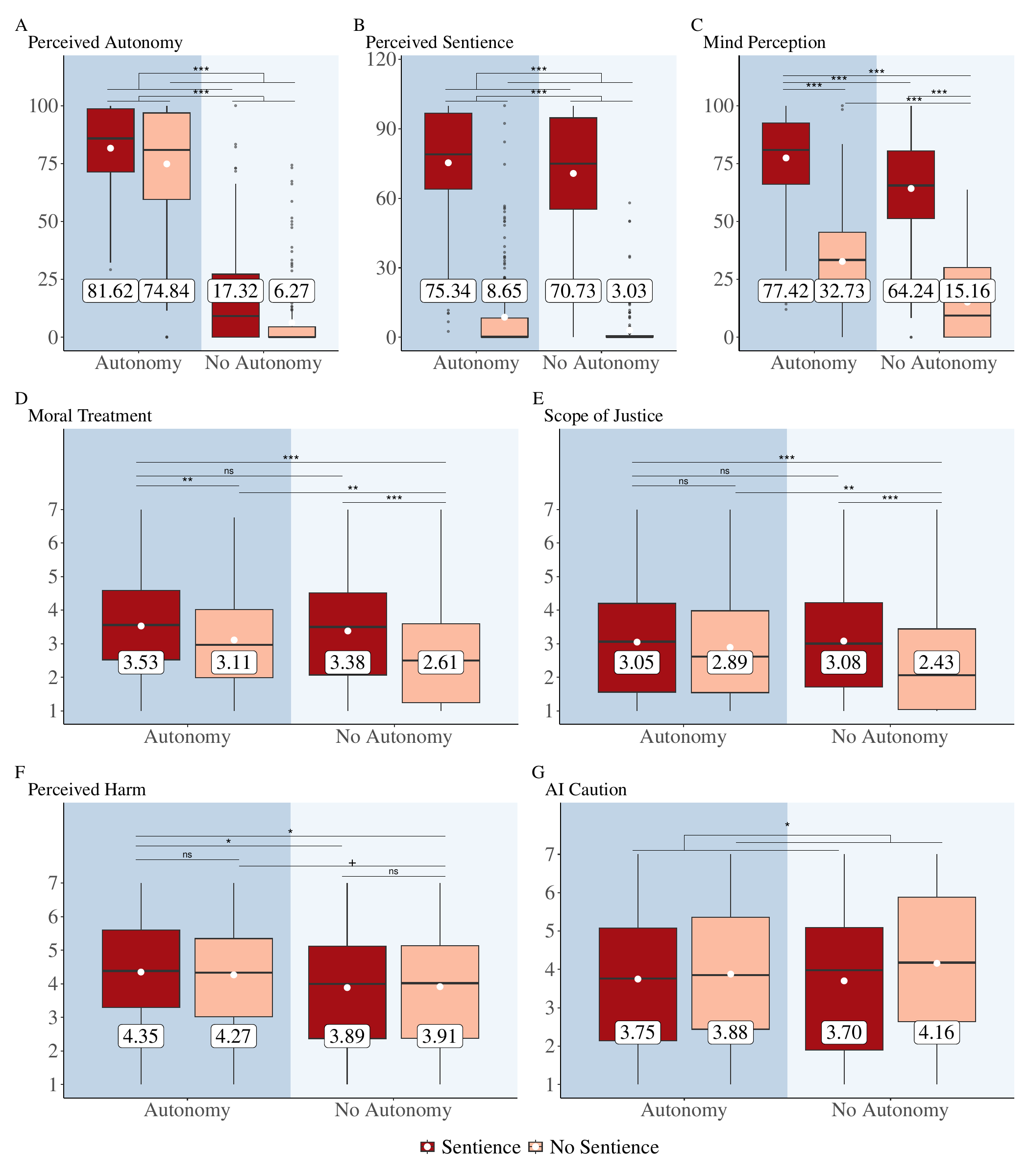}
    \caption{Autonomy and sentience effects (Experiment 3). Note. \textit{ns} = non-significant, +$p$ < .10, *$p$ < .05, **$p$ < .01, ***$p$ < .001. Red shows the sentience main effect (dark = sentience, light = no sentience). Blue backgrounds the autonomy main effect (dark = autonomy, light = no autonomy). The H1 and H2 significance tests are shown for Perceived Autonomy and Perceived Sentience. The H3 and H4 pairwise comparisons are shown for Mind Perception, Perceived Harm, Moral Treatment, and Scope of Justice. The significant sentience effect is shown for AI Caution. The means are labeled over each box and displayed as a white point. The black midpoint line is the median. The edges of the box are the lower and upper quartile values. The whiskers are 1.5*IQR, the default in R’s ggplot2 package. Dots are outliers.}
    \label{fig:figure5}
    \Description{This multi-panel boxplot shows the average responses and distributions for the dependent variables in Experiment 3. The x-axis of each boxplot panel shows the autonomy and no autonomy experimental conditions. The sentience and no sentience experimental conditions are subdivided under autonomy and no autonomy. The y-axes show the measurement scale for the dependent variables. Each boxplot panel shows the mean, median, quartile values, inter-quartile range, and any outliers. Row 1 shows a significantly higher mean for autonomy than no autonomy, and sentience than no sentience on (A) Perceived Autonomy, (B) Perceived Sentience, and (C) Mind Perception. For (C) Mind Perception, the combined autonomy and sentience condition is significantly higher than the other conditions and the combined no autonomy and no sentience condition is significantly lower than the other conditions. Row 2 (D) Moral Treatment shows a significantly higher mean for the combined autonomy and sentience condition than the combined autonomy and no sentience condition and the combined no autonomy and no sentience condition, which also has a significantly lower mean than the combined sentience and no autonomy condition and the combined autonomy and no sentience condition. The same pattern is shown in Row 2 (E) Scope of Justice, but without the significant difference between the combined autonomy and sentience condition and the combined autonomy and no sentience condition. Row 3 (F) Perceived Harm shows a significantly higher mean for the combined autonomy and sentience condition than the combined no autonomy and no sentience condition and the combined sentience and no autonomy condition. Row 3 (G) AI Caution shows a significantly higher mean for the no sentience than sentience conditions on (G) AI Caution.}    
\end{figure*}

Autonomy relative to no autonomy, increased Perceived Autonomy, $F$(1, 737) = 2034.02, $p$ < .001, $\eta_p^2$ = .73. Sentience also increased Perceived Autonomy relative to no sentience, $F$(1, 737) = 36.62, $p$ < .001, $\eta_p^2$ = .05. There was no interaction, $F$(1, 737) = 2.09, $p$ = .148, $\eta_p^2$ < .01.

Sentience increased Perceived Sentience relative to no sentience, $F$(1, 737) = 2074.53, $p$ < .001, $\eta_p^2$ = .74. Autonomy increased Perceived Sentience, $F$(1, 737) = 12.00, $p$ = .001, $\eta_p^2$ = .02, relative to no autonomy. There was no interaction, $F$(1, 737) = 0.12, $p$ = .734, $\eta_p^2$ < .01. 	

Mind Perception increased with autonomy relative to no autonomy, $F$(1, 737) = 122.71, $p$ < .001, $\eta_p^2$ = .14, and sentience relative to no sentience, $F$(1, 737) = 1140.84, $p$ < .001, $\eta_p^2$ = .61, consistent with Experiments 1 and 2. The interaction between autonomy and sentience was not significant, $F$(1, 737) = 2.50, $p$ = .114, $\eta_p^2$ < .01, although the hypothesized H3 and H4 patterns emerged in false discovery rate (FDR) corrected post hoc pairwise comparisons. Mind Perception was significantly lower for participants in the None condition, and significantly higher for participants in the Both condition ($p$s < .001; see \Cref{fig:figure5} and Supplementary Table S6 for descriptive statistics).

Autonomy relative to no autonomy significantly increased Moral Treatment, $F$(1, 737) = 9.22, $p$ = .002, $\eta_p^2$ = .01, and increased Scope of Justice, $F$(1, 737) = 3.54, $p$ = .060, $\eta_p^2$ = .01, although not statistically significantly. Sentience relative to no sentience increased Moral Treatment, $F$(1, 737) = 30.91, $p$ < .001, $\eta_p^2$ = .04, and Scope of Justice, $F$(1, 737) = 12.49, $p$ < .001, $\eta_p^2$ = .02. The interaction between autonomy and sentience on Moral Treatment was not significant, $F$(1, 737) = 2.72, $p$ = .100, $\eta_p^2$ < .01 although FDR-corrected post hoc pairwise comparisons supported H4b and partly supported H3b. Moral Treatment was significantly lower in the None condition than in the other conditions ($p$s < .01), and significantly higher in the Both condition than in the Autonomy (i.e., autonomous but not sentient) condition ($p$ = .009). There was no statistical difference between the Both and Sentience (i.e., sentient but not autonomous) conditions ($p$ = .336). The interaction was significant on Scope of Justice, $F$(1, 737) = 4.68, $p$ = .031, $\eta_p^2$ = .01, supporting H4b but not H3b. Scope of Justice was significantly lower in the None condition ($p$s < .01).

Autonomy relative to no autonomy increased Perceived Harm, $F$(1, 737) = 10.44, $p$ = .001, $\eta_p^2$ = .01, and had no effect on AI Caution, $F$(1, 737) = 0.71, $p$ = .401, $\eta_p^2$ < .01, consistent with Experiment 1. Sentience did not change Perceived Harm relative to no sentience, $F$(1, 737) = 0.05, $p$ = .822, $\eta_p^2$ < .01, but there was a significant effect on AI Caution, $F$(1, 737) = 4.39, $p$ = .037, $\eta_p^2$ = .01. Sentience decreased AI Caution relative to no sentience, contrary to H2c and the Experiment 2 results. The interaction between autonomy and sentience was not significant for either Perceived Harm, $F$(1, 737) = 0.20, $p$ = .659, $\eta_p^2$ < .01, or AI Caution, $F$(1, 737) = 1.38, $p$ = .241, $\eta_p^2$ < .01. FDR-corrected comparisons showed that the Both condition was significantly higher on Perceived Harm than the None and Sentience conditions ($p$s = .049), partly supporting H3c. The None condition was lower than the Autonomy condition, although not statistically significantly ($p$ = .067). There were no significant comparisons for AI Caution. Unexpectedly, the effect of AI sentience on Perceived Harm disappeared and the effect on AI Caution flipped from Experiment 2, such that sentience decreased AI Caution. These differences could reflect the different news media context (i.e., post-ChatGPT) in which Experiment 3 was conducted. Alternatively, the differences might be due to the explicitly labeled disentanglement of sentience from autonomy, especially if this disentanglement contradicts pre-existing mental models of sentience that embed autonomy as a part of sentience. We address this in Experiment 4. 

\section{Experiment 4}

We removed the explicit conceptual labels for autonomy and sentience used in Experiments 1-3 to reduce any potential dissonance created by our disentanglement and users' pre-existing mental models, which may have been activated merely by the use of labels in Experiments 1-3. We also approximately doubled the sample size to increase statistical power given the underpowered interaction effects in Experiment 3. Experiment 4 was preregistered (\url{https://osf.io/s698b}).

\subsection{Methods}

The materials, open data, and R analysis code are available on the OSF (\url{https://osf.io/mquvw/}). 

\subsubsection{Participants and Design}

Participants from Experiments 1-3 were excluded from recruitment. We recruited 1,530 U.S. adults from Prolific. Participants who failed the attention check (None $n$ = 14, Autonomy $n$ = 9, Sentience $n$ = 22, Both $n$ = 21) were excluded from analyses\footnote{Like in Experiment 3, we included attention check responses > 6.90 ($n$ = 77).} and 13 additional participants were excluded due to an unknown technical error leading to no recorded data. The final sample was 1,451 with random assignment to autonomy ($n$ = 726) or no autonomy ($n$ = 725) and sentience ($n$ = 711) or no sentience ($n$ = 740) information (four fully-crossed conditions: None $n$ = 382, Autonomy $n$ = 358, Sentience $n$ = 343, Both $n$ = 368). The sample was 49\% women ($M_{age}$ = 41.60, \textit{SD} = 14.06, range = 18-98) and 72\% White, 11\% Black, 7\% Latin or Hispanic, 6\% Asian, and 4\% multiracial, indigenous, or other. Participants self-identified as somewhat liberal (\textit{M} = 2.54, \textit{SD} = 1.18, range = 1-5), most owned smartphones (85\%), some owned robotic or AI devices (47\%), and some worked with AI (24\%). Participants self-reported directly interacting with AI occasionally (\textit{M} = 2.11, \textit{SD} = 1.93, range = 0-5) and being occasionally exposed to AI narratives (\textit{M} = 2.37, \textit{SD} = 1.42, range = 0-5). 

\subsubsection{Procedure}

The Experiment 4 procedure replicated Experiment 3 except for the modification to remove the explicit autonomy and sentience labels.

\paragraph{Measures}

The dependent variables matched the previous experiments and were reliable (Perceived Autonomy: Cronbach’s $\alpha$ = .98; Perceived Sentience: Cronbach’s $\alpha$ = .98; Mind Perception: Cronbach’s $\alpha$ = .89; Moral Treatment: Cronbach’s $\alpha$ = .87; Scope of Justice: Cronbach’s $\alpha$ = .91; Perceived Harm: Cronbach’s $\alpha$ = .94; AI Caution: Cronbach’s $\alpha$ = .91). The same exploratory questions and demographics were measured as in Experiment 3. 

\subsection{Results}

\begin{figure*}[htbp]
    \centering
    \includegraphics[width=0.9\textwidth, height=0.85\textheight, keepaspectratio]{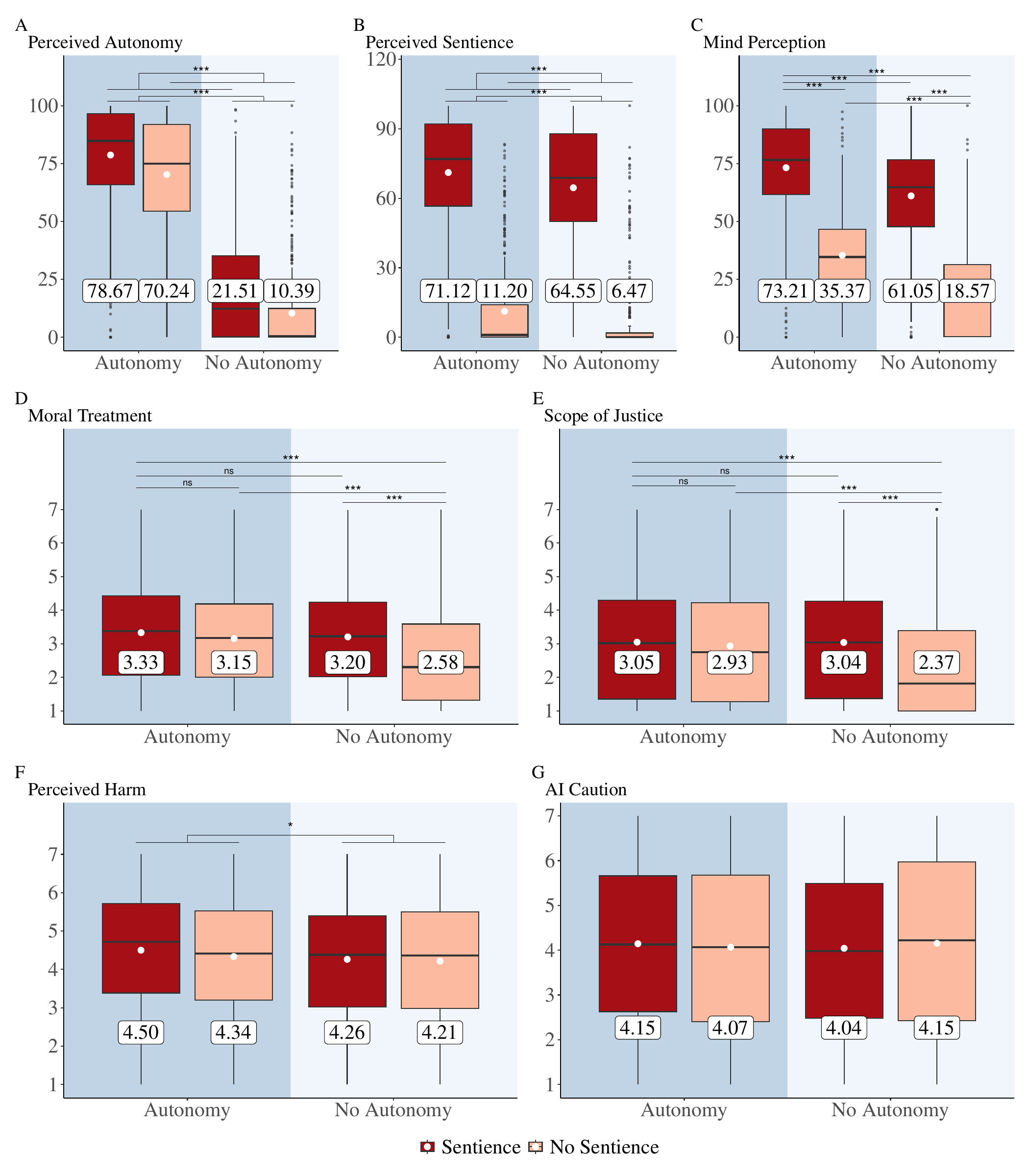}
    \caption{Robust autonomy and sentience effects (Experiment 4). Note. \textit{ns} = non-significant, +$p$ < .10, *$p$ < .05, **$p$ < .01, ***$p$ < .001. Red shows the sentience main effect (dark = sentience, light = no sentience). Blue backgrounds the autonomy main effect (dark = autonomy, light = no autonomy). The H1 and H2 significance tests are shown for Perceived Autonomy and Perceived Sentience. The H3 and H4 pairwise comparisons are shown for Mind Perception, Moral Treatment, and Scope of Justice. The significant autonomy effect is shown for Perceived Harm. There were no significant effects on AI Caution. The means are labeled over each box and displayed as a white point. The black midpoint line is the median. The edges of the box are the lower and upper quartile values. The whiskers are 1.5*IQR, the default in R’s ggplot2 package. Dots are outliers.}
    \label{fig:figure6}
    \Description{This multi-panel boxplot shows the average responses and distributions for the dependent variables in Experiment 4. The x-axis of each boxplot panel shows the autonomy and no autonomy experimental conditions. The sentience and no sentience experimental conditions are subdivided under autonomy and no autonomy. The y-axes show the measurement scale for the dependent variables. Each boxplot panel shows the mean, median, quartile values, inter-quartile range, and any outliers. Row 1 shows a significantly higher mean for autonomy than no autonomy, and sentience than no sentience on (A) Perceived Autonomy, (B) Perceived Sentience, and (C) Mind Perception. For (C) Mind Perception, the combined autonomy and sentience condition is significantly higher than the other conditions and the combined no autonomy and no sentience condition is significantly lower than the other conditions. Row 2 (D) Moral Treatment shows a significantly higher mean for the combined autonomy and sentience condition than the combined no autonomy and no sentience condition, which also has a significantly lower mean than the combined sentience and no autonomy condition and the combined autonomy and no sentience condition. The same pattern is shown in Row 2 (E) Scope of Justice. Row 3 shows a significantly higher mean for autonomy than no autonomy on (F) Perceived Harm and no significant differences on (G) AI Caution.}
\end{figure*}

Autonomy relative to no autonomy increased Perceived Autonomy, $F$(1, 1447) = 2345.12, $p$ < .001, $\eta_p^2$ = .62. Sentience relative to no sentience increased Perceived Autonomy, $F$(1, 1447) = 65.42, $p$ < .001, $\eta_p^2$ = .04. There was no interaction, $F$(1, 1447) = 1.24, $p$ = .266, $\eta_p^2$ < .01.

Sentience increased Perceived Sentience relative to no sentience, $F$(1, 1447) = 2352.26, $p$ < .001, $\eta_p^2$ = .62. Autonomy increased Perceived Sentience relative to no autonomy, $F$(1, 1447) = 21.57, $p$ < .001, $\eta_p^2$ = .02. There was no interaction, $F$(1, 1447) = 0.57, $p$ = .450, $\eta_p^2$ < .01. 

Mind Perception significantly increased for autonomy relative to no autonomy, $F$(1, 1447) = 181.72, $p$ < .001, $\eta_p^2$ = .11, and sentience relative to no sentience, $F$(1, 1447) = 1397.51, $p$ < .001, $\eta_p^2$ = .49. The interaction between autonomy and sentience was significant, $F$(1, 1447) = 4.66, $p$ = .031, $\eta_p^2$ = .003 (see \Cref{fig:figure6}). FDR-corrected post hoc pairwise comparisons supported H3a and H4a. All combinations significantly differed from each other ($p$s < .001; see \Cref{fig:figure6} and Supplementary Table S7 for descriptive statistics).

Moral Treatment significantly increased for autonomy relative to no autonomy, $F$(1, 1447) = 21.40, $p$ < .001, $\eta_p^2$ = .02, as did Scope of Justice, $F$(1, 1447) = 11.38, $p$ < .001, $\eta_p^2$ = .01, consistent with Experiments 1 and 3. Sentience relative to no sentience increased Moral Treatment, $F$(1, 1447) = 27.98, $p$ < .001, $\eta_p^2$ = .02, and Scope of Justice, $F$(1, 1447) = 21.31, $p$ < .001, $\eta_p^2$ = .02, consistent with Experiments 2 and 3. Autonomy and sentience significantly interacted on Moral Treatment, $F$(1, 1447) = 8.68, $p$ = .003, $\eta_p^2$ = .01. Supporting H4b, Moral Treatment was significantly lower for the None mental model ($p$s < .001). Autonomy and sentience also significantly interacted on Scope of Justice, $F$(1, 1447) = 10.65, $p$ = .001, $\eta_p^2$ = .01. Scope of Justice was significantly lower for the None mental model ($p$s < .001), supporting H4b. H3b was not supported for either Moral Treatment or Scope of Justice.

Autonomy relative to no autonomy increased Perceived Harm, $F$(1, 1447) = 4.10, $p$ = .043, $\eta_p^2$ = .003, but not AI Caution, $F$(1, 1447) = 0.01, $p$ = .936, $\eta_p^2$ < .01, consistent with Experiments 1 and 3. Sentience relative to no sentience did not change Perceived Harm, $F$(1, 1447) = 1.44, $p$ = .230, $\eta_p^2$ < .01, consistent with Experiment 3. There was no effect of sentience on AI Caution, $F$(1, 1447) = 0.03, $p$ = .869, $\eta_p^2$ < .01, a third pattern of results from Experiments 2 and 3. Autonomy and sentience did not significantly interact on either Perceived Harm, $F$(1, 1447) = 0.43, $p$ = .513, $\eta_p^2$ < .01, or AI Caution, $F$(1, 1447) = 0.86, $p$ = .355, $\eta_p^2$ < .01. There were no significant post hoc pairwise comparisons on Perceived Harm or AI Caution (see \Cref{fig:figure6}). H3c and H4c were not supported. Exploratory path analyses suggested a mechanistic role for perceptions of autonomy and sentience, consistent with theories linking the perception of mental capacities to moral consideration \cite{gray_morality_2012} (see Supplementary Figures S5-12).

With a larger sample size, the significant interaction between autonomy and sentience appeared for Mind Perception, Moral Treatment, and Scope of Justice. Perceived Harm was still only increased by autonomy, which again had a null effect on AI Caution. The null effect of sentience on Perceived Harm matched Experiment 3, and the sentience pattern on AI Caution changed to a null effect. Experiment 4 results largely cohered with the previous experiments, suggesting that the effects of autonomy and sentience are robust to explicit conceptual labels that might have sparked dissonance with pre-existing mental models. Across experiments, the effects of sentience appeared larger than autonomy, suggesting that mental models of sentience might be more impactful than mental models of autonomy on reactions to AI. We examine this with a meta-analysis of the data from Experiments 1-4. 

\section{Meta-Analysis}

\begin{figure*}[htbp]
    \centering
    \includegraphics[width=0.6\linewidth]{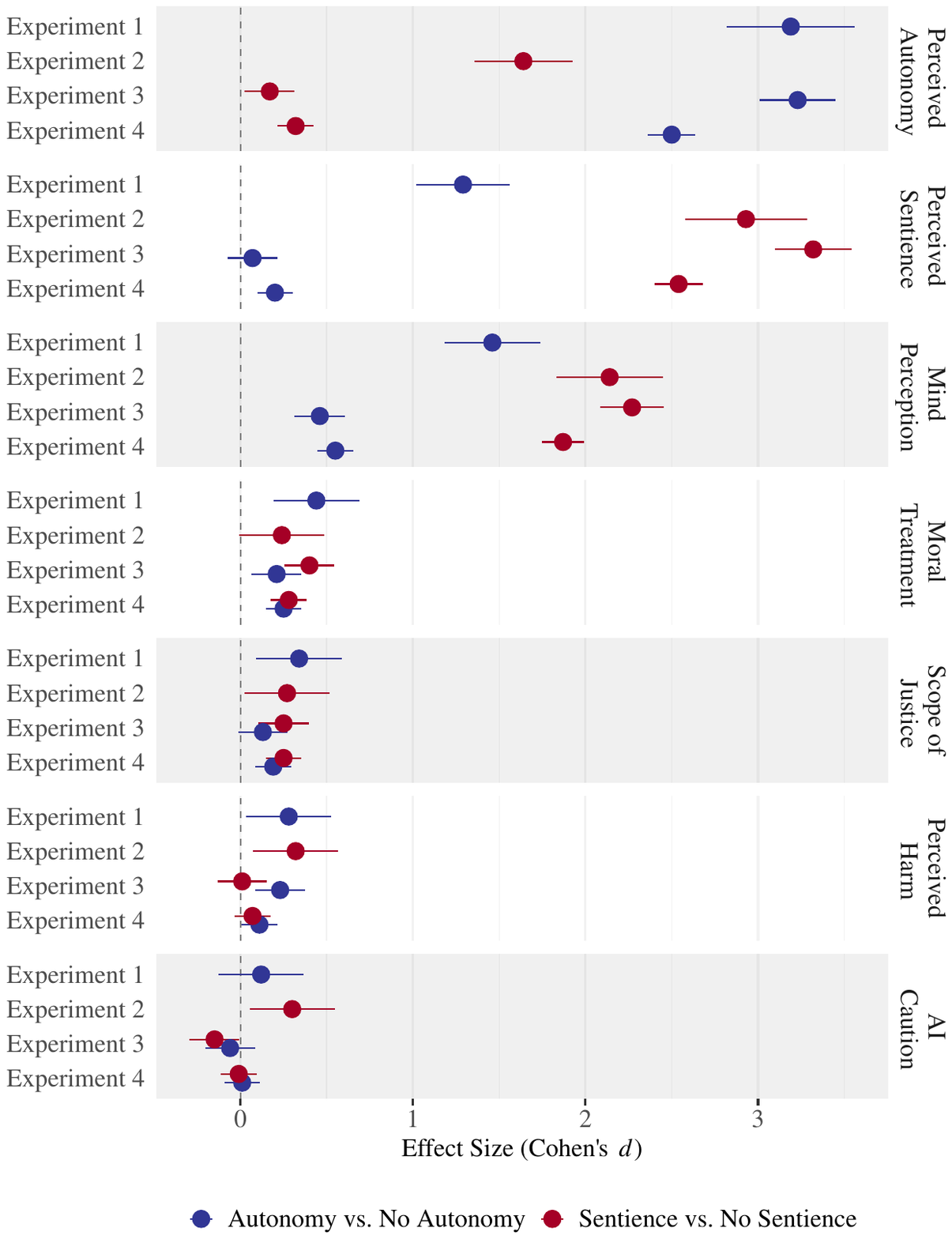}
    \caption{Cohen's d effect sizes across four experiments and seven outcomes. Note. The error bars show the \textit{95\% CI}. See Supplementary Materials for the Cohen’s $d$ values. }
    \label{fig:figure7}
    \Description{This forest plot shows the Cohen's d effect size values for the autonomy and sentience experimental conditions on each dependent variable subdivided by experiment. The effect of autonomy compared to no autonomy is shown with a blue dot and confidence interval whiskers. The effect of sentience compared to no sentience is shown with a red dot and confidence interval whiskers. Row 1 shows autonomy and sentience effect sizes on Perceived Autonomy for each experiment. Autonomy has a larger effect in Experiments 1, 3, and 4 than sentience has in Experiments 2, 3, and 4. Row 2 shows autonomy and sentience effect sizes on Perceived Sentience for each experiment. Sentience has a larger effect in Experiments 2, 3, and 4 than autonomy has in Experiments 1, 3, and 4. Comparing Rows 1 and 2, the sentience effects on Perceived Autonomy are larger than the autonomy effects on Perceived Sentience. Row 3 shows autonomy and sentience effects on Mind Perception for each experiment. Sentience has a larger effect in Experiments 2, 3, and 4 than autonomy has in Experiments 1, 3, and 4. Row 4 and 5 show autonomy and sentience effects on Moral Treatment and Scope of Justice, respectively, for each experiment. Sentience effects tended to be larger than autonomy effects in Experiments 3 and 4, whereas the autonomy effect appears slightly larger in Experiment 1 than the sentience effect in Experiment 2. Row 6 shows autonomy and sentience effects on Perceived Harm for each experiment. Autonomy effects are significant in Experiments 1, 3, and 4. The sentience effect is significant in Experiment 2 but not in Experiment 3 or 4. Row 7 shows autonomy and sentience effects on AI Caution for each experiment. Autonomy effects are non-significant in Experiments 1, 3, and 4. The sentience effect is significantly positive in Experiment 2, significantly negative in Experiment 3, and null in Experiment 4.}
\end{figure*}

We conducted a meta-analysis \cite{goh_mini_2016} across the four experiments to examine whether there is an overall difference in magnitude between autonomy and sentience effects given that sentience appeared to have larger effects than autonomy.\footnote{The meta-analysis R code is on the OSF: \url{https://osf.io/fbt3r/}.} We tested this statistically by comparing Cohen’s \textit{d} effect sizes (see \Cref{fig:figure7}) using multivariate restricted maximum likelihood mixed effects. Autonomy and sentience were fixed effects. Study and dependent variable were random effects. First, we established that there are significant meta-analytic effects across dependent variables for autonomy relative to no autonomy, Cohen’s $d$ = 0.72, \textit{SE} = 0.22, $z$ = 3.25, $p$ < .001, \textit{95\% CI} [0.29, 1.15], and sentience relative to no sentience, Cohen’s $d$ = 0.92, \textit{SE} = 0.24, $z$ = 3.80, $p$ < .001, \textit{95\% CI} [0.45, 1.40]. Second, we tested whether sentience had a larger effect than autonomy on the 42 effect sizes from our four studies. Sentience, average Cohen’s $d$ = 0.90, \textit{SE} = .16, $z$ = 5.67, $p$ < .001, \textit{95\% CI} [0.59, 1.22], was more impactful than autonomy, average Cohen’s $d$ = 0.70, \textit{SE} = .16, $z$ = 4.37, $p$ < .001, \textit{95\% CI} [0.38, 1.01], on reactions, $QM$(2) = 98.20, $p$ < .001, \textit{95\% CI} [-0.26, -0.16]. A sensitivity analysis excluding Perceived Autonomy and Perceived Sentience showed consistent results (see Supplementary Materials). 

\section{General Discussion}

AI is transforming our economic, social, and moral experiences \cite{anderson_visions_2022, matthews_evolution_2021, rahwan_machine_2019, ritchie_technology_2017, russell_human_2019, tegmark_life_2017}. Relationships with AIs like Replika and ChatGPT are becoming more common as people seek companionship \cite{pentina_exploring_2023}, mental health support \cite{de_freitas_ai_2025, giray_cases_2025, maples_loneliness_2024}, and efficient assistance \cite{skjuve_user_2023}. Interactions with such companions and agents can co-activate mental models of sentience and autonomy that shape reactions to AI systems. We observed three types of reactions to activated mental models of autonomy and sentience in four experiments, supporting H1a-c, H2a and H2b, H3a, and H4a and H4b. H2c was supported in Experiment 2. H3b and H3c were partly supported in Experiment 3, and H4c was not supported (see \Cref{tab:hypothesis_summary} and \Cref{fig:figure1}). Activating mental models of autonomy and sentience increased the perception of autonomy, sentience, and mind in one AI, as well as moral consideration for AI in general. Autonomy increased the perceived threat of AI. The co-activation of autonomy and sentience produced the most mind perception. The co-activation of no autonomy and no sentience produced minimal mind perception and moral consideration.

\begin{table*}[htbp]
\centering
\caption{Summary of evidence for hypotheses}
\label{tab:hypothesis_summary}
\begin{tabular}{p{0.05\textwidth}p{0.50\textwidth}p{0.05\textwidth}p{0.05\textwidth}p{0.05\textwidth}p{0.05\textwidth}}
\hline
\multicolumn{2}{l}{} & \multicolumn{4}{c}{Supported} \\
\multicolumn{2}{l}{Hypothesis} & E1 & E2 & E3 & E4 \\
\hline
\multicolumn{2}{l}{Autonomy hypotheses} \\
\textbf{H1a} & AI autonomy increases the perception of mind. & Yes & {} & Yes & Yes \\
\textbf{H1b} & AI autonomy increases the moral consideration of AIs. & Yes & {} & Yes & Yes \\
\textbf{H1c} & AI autonomy increases the perceived threat of AIs. & Yes & {} & Yes & Yes \\
\\
\multicolumn{2}{l}{Sentience hypotheses} \\
\textbf{H2a} & AI sentience increases the perception of mind. & {} & Yes & Yes & Yes \\
\textbf{H2b} & AI sentience increases the moral consideration of AIs. & {} & Yes & Yes & Yes \\
\textbf{H2c} & AI sentience increases the perceived threat of AIs. & {} & Yes & No & No \\
\\
\multicolumn{2}{l}{Combined autonomy and sentience hypotheses} \\
\textbf{H3a} & Combined autonomy and sentience produce the most perception of mind. & {} & {} & Yes & Yes \\
\textbf{H3b} & Combined autonomy and sentience produce the most moral consideration. & {} & {} & Partial & No \\
\textbf{H3c} & Combined autonomy and sentience produce the most perceived threat. & {} & {} & Partial & No \\
\\
\multicolumn{2}{l}{Combined no autonomy and no sentience hypotheses} \\
\textbf{H4a} & Combined no autonomy and no sentience produces the least perception of mind. & {} & {} & Yes & Yes \\
\textbf{H4b} & Combined no autonomy and no sentience produces the least moral consideration. & {} & {} & Yes & Yes \\
\textbf{H4c} & Combined no autonomy and no sentience produces the least perceived threat. & {} & {} & No & No \\
\hline
\end{tabular}
\end{table*}

\subsection{Sentience as a Foundational Mental Model}

Overall, sentience had a larger effect than autonomy on mind perception, aligning with evidence that sentience underpins mind perception \cite{gray_dimensions_2007, gray_feeling_2012, koban_it_2024, weisman_reasoning_2015}. Sentience also had a larger effect on perceived autonomy than autonomy had on perceived sentience. Consistent with previous research \cite{chernyak_childrens_2016, opfer_identifying_2002, weisman_reasoning_2015}, sentience and autonomy are conceptually connected, and our results suggest that mental models of sentience activate, and may anchor, mental models of autonomy \cite{weisman_reasoning_2015}. Sentience is also intertwined with other concepts (e.g., life, meta-cognition). This network of mental models may explain the variable effects we observed of sentience on perceived harm and policy preferences. A mental model of sentience that does not disentangle it from autonomy, such as in Experiment 2, might increase perceived threat because sentience automatically co-activates a mental model of autonomy, a supposition supported by the Experiment 4 exploratory analyses (see Supplementary Figures S5-12). 

Fearful reactions to AI sentience may fluctuate following changes in the scientific, news, or science fiction narratives circulating about AI \cite{banks_optimus_2020, gilardi_we_2024}. AI narratives activate many mental models alongside sentience, such as sapience, which can threaten perceptions of humanity’s unique identity. Pioneering research on human essentialism showed that capacities typically considered to be uniquely human (e.g., sapience, autonomy) are more threatened by nonhumans than are capacities (e.g., pain, sensory perception) typically considered to be overlapping with nonhumans (e.g., chimpanzees, chickens) \cite{haslam_dehumanization_2014}. This suggests that AI narratives that activate sentience alongside sapience or autonomy may increase perceived threat more than narratives emphasizing perception or affect. Despite this complexity, sentience increased mind perception and moral consideration, aligning with HCI and psychological research on experiential capacities \cite{nijssen_saving_2019, sommer_childrens_2019}. These results have implications for deepening explanations of human-AI interaction beyond human-likeness and understanding how people form lasting relationships with AIs. This is relevant now, as people increasingly interact with chatbots like ChatGPT and Replika that recognize user emotions, express emotions, and appear to take independent actions.

\subsection{Autonomy as a Source of Threat and Motivation for Governance}

Autonomy drove the perception of AI harm, a facet of perceived threat, consistent with previous research \cite{zlotowski_can_2017}. Although Perceived Harm and AI Caution were significantly, strongly correlated\footnote{We base our judgment of effect size strength on Schäfer and Schwarz's (2019) demonstration of effect sizes in preregistered psychology studies.} across experiments (Experiment 1: \textit{r} = .51, Experiment 2: \textit{r} = .49, Experiment 3: \textit{r} = .45, Experiment 4: \textit{r} = .58, see Supplementary Figures S1-4), autonomy did not increase policy support. Further, our exploratory analyses in Experiment 4 found some preliminary evidence supporting mental model activation effects on the perception of harm, but not policy support, through increased perceptions of autonomy (see Supplementary Figures S5-12). Mental models of one AI’s capacities may have little effect on policy preferences, although information about one AI’s actions have been shown to spill over to judgments of AI in general \cite{longoni_algorithmic_2023, manoli_ai_2025}. Spillover effects might explain how people form mental models of AI as a monolithic technology from one exemplar, akin to human intergroup transfer effects and attitude generalization processes \cite{tausch_secondary_2010, ware_cognitive_1967} that shape the perception of a whole group of people from an interaction with one group member. The more we recognize the heterogeneity of AI capacities and roles, and users' mental models, the less we may misperceive, misattribute, and overgeneralize from a generic mental model of “AI,” to all AI systems, or from the features or actions of one AI to all AIs. Disentangling the effects of mental models on reactions to AI may also contribute to AI literacy initiatives and reducing risks associated with misunderstanding AI, such as unwarranted trust \cite{malle_trust_2020}, as users become more familiar with their mental models about, and expectations for, various AI systems' behaviors. 

Perceived threats can be construed as realistic and symbolic following Integrated Threat Theory \cite{stephan_intergroup_2015}. AI can be perceived as a tangible threat (i.e., realistic) to resources like jobs, energy, and human existence. People can also perceive intangible threats (i.e., symbolic) to human uniqueness, values, and autonomy. Perceiving AI to be harmful and supporting policies banning advanced AI developments, as we measured them, collapse across realistic and symbolic threats. Given that people perceive AIs as an outgroup \cite{smith_human-robot_2021, vanman_danger_2019}, distinguishing between these types of threats may be critical to expand our understanding of the multidimensionality of AI threat \citep[e.g.,][]{bozkurt_artificial_2023, kieslich_threats_2021, li_dimensions_2020, zhan_what_2023} and the triggers of threat responses within AI design and policymaking.

\subsection{Mind Perception, Deconstructed}

We used an established scale to measure mind perception, and we adapted that scale to measure perceptions of autonomy and sentience. These three measures are conceptually related, and differentiating them touches on profound questions about what mind is, whether there are dimensions of mind, who has a mind, and how perceptions of mind impact downstream attitudes and behaviors. Although our studies cannot meaningfully contribute to the metaphysics of these questions, they offer evidence regarding mind perception dimensionality and downstream impacts. Studies of mind perception typically describe mind as one dimension (i.e., “mind”), two dimensions (i.e., experience, agency), or three dimensions (i.e., experience, cognition, planning) \citep[e.g.,][]{kozak_what_2006, gray_dimensions_2007, tzelios_evidence_2022}. 

We observed that activating mental models of autonomy and sentience strongly affected the established mind perception measure, as expected. The Mind Perception, Perceived Autonomy, and Perceived Sentience measures were positively correlated (\textit{r} = .16-.89; see Supplementary Figures S1-4). The strongest correlations were between Mind Perception and Perceived Sentience (Experiment 1: \textit{r} = .84, Experiment 2: \textit{r} = .89, Experiment 3: \textit{r} = .87, Experiment 4: \textit{r} = .87), consistent with research arguing that perceived sentience underpins mind perception \cite{koban_it_2024, weisman_reasoning_2015}. We observed the strongest correlations between Mind Perception and Perceived Autonomy (\textit{r} = .82), Mind Perception and Perceived Sentience (\textit{r} = .89), and Perceived Autonomy and Perceived Sentience (\textit{r} = .78) in Experiment 2, in which the sentience or no sentience mental model was activated. These correlations again support the idea that mental models of sentience have a strong influence on mind perception.

We further examined the operational separability of the measurements with exploratory factor analyses (see Supplementary Tables S10-25). We observed a three-dimensional structure, consistent with previous three-dimensional evidence \cite{kozak_what_2006, malle_how_2019}. In these analyses, sentience is distinguishable from autonomy, and sentience explains the most variation in mind perception, consistent with evidence from \citet{koban_it_2024} and \citet{weisman_reasoning_2015}. This exploratory analysis suggests that the conceptual and empirical relationships between these perceptions require additional study, particularly to account for sentience as not necessarily reducible to the more often studied, “experience,” which may be a broader construct that includes embodiment, emotion, and assumptions based on biological evolutionary processes. Expanding the study of sentience in HCI is especially important as human-AI interaction occurs with increasingly powerful AI systems that are seen as having mind and sentience \cite{anthis_perceptions_2025}.

\subsection{Design Implications}

Designers should consider that activating mental models of specific capacities may trigger distinct responses. People react differently to AI systems that present as more autonomous (e.g., autonomous vehicles, web browser agents) or more sentient (e.g., companion AI, emotion AI) because of the activation of their corresponding mental model. For instance, systems that are designed to voluntarily switch between roles as an assistant and friend during conversations could activate a mental model of autonomy and increase the perception of AI autonomy. Systems designed to use personal pronouns to express empathy or compassion towards a user in human-like language could activate a mental model of sentience and increase the perception of AI sentience. Additionally, features like this, that cue social and mind perception, may co-activate other mental models such as life and sapience.

This supports the judicious use of targeted tuning strategies \cite{inkpen_advancing_2023}, such as increasing or decreasing anthropomorphism by increasing or decreasing a chatbot’s use of personal pronouns and human-like conversational conventions, or instructing users to dehumanize AI systems during interactions \cite{bi_i_2023, maeda_when_2024, yam_reducing_2021}. Designs aimed at tuning users’ perceptions of autonomy and sentience must avoid misleading users about the system’s actual capacities. Doing so could increase the risk of harm. Tuning perceptions of autonomy could lead to unwarranted trust in an AI’s advice, which may have consequences for mental health safety \cite{sturgeon_humanagencybench_2025}, increased rejection of AI due to heightened risk perception \citep[e.g.,][]{bullock_public_2025}, and human-AI agent cooperation \cite{duan_trusting_2025, lee_toward_2025}. Questions of mental health safety and multicultural therapy efficacy have already been posed with ChatGPT use \cite{aleem_towards_2024, giray_cases_2025, jung_ive_2025, suriano_theory_2025}. Tuning perceptions of sentience could lead to the mistreatment of the AI, behaviors that could transfer to antisocial human-human and human-animal relations \citep[e.g.,][]{caviola_what_2025, coghlan_could_2019, guingrich_ascribing_2024, kanepajs_what_2025, reese_end_2018, wilks_why_2026}, and increased emotional attachments to an AI that endanger users \cite{knox_harmful_2025}. Such consequences have already been observed with user susceptibility to chatbot manipulation \cite{ho_potential_2025} and users seeking help for negative body image confused by widely used chatbots (i.e., Replika and Wysa) \cite{an_field_2023}.

Further, designers should consider differences in user experiences if AI systems appear more autonomous or more sentient. Cues signaling autonomy or agency trigger perceptions of sentience, regardless of the system's actual capacities. This may lead to users forming attachments and emotional bonds with AIs, including those not designed for companionship, that affect the quality of their interactions. Likewise, cues signaling sentience increase perceptions of autonomy that in turn increase feelings of being threatened. Co-activations of autonomy and sentience may increase experiences of uncanny emotions that could lead to a rejection of AI in general \cite{kitchens_fearful_2025, manoli_ai_2025}, something requiring future study. 

Finally, mapping the content of mental models, understanding the cues that activate them, and the impacts of mental model activations on downstream thoughts and behaviors can bridge gaps between AI designers and users. \citet{xie_how_2017} found that different methods of activating mental models revealed different types of similarity between designer and user mental models. To better align designer and user mental models, future research is needed to map the content of generic and specific mental models of AI systems, contexts, and capacities like autonomy and sentience, something our pilot studies begin to do. Future experiments building on our experiments are needed to understand the effects of activating mental models, and the maintenance and updating of mental models under exposure to novel information \citep[e.g.,][]{vandenbosch_information_1996}. Mental models may update more often given ongoing interactions with constantly evolving AI agents such as ChatGPT.

\subsection{Policy Implications}

The current results also inform the regulation of AI development and social integration. AI literacy initiatives to educate the public on the multifaceted capacities of AI systems can help to increase perceptions of AI heterogeneity and challenge misperceptions of AI as a monolithic technology formed from exposure to AI narratives in cultural products (e.g., news media). Increased AI literacy could bound complacency. Complacency can lead to outcomes such as unwarranted trust in AI recommendations, or, on the other hand, mass rejection of AI systems based on a generic misperception. It could also lead to more appropriate, domain-specific uses of systems such as Wysa for mental health support rather than ChatGPT or Replika. Reducing monolithic perceptions of AI could increase the adoption of domain-specific tools (e.g., medical decision-making systems, agent AIs that plan and organize events at work), and increasing perceptions of AI heterogeneity could mitigate the extent to which negative attitudes towards one AI generalize to all AIs. 

Similarly, sensitivity training policies to educate designers on the complex effects of user mental models could lead to finetuned designs and better user experiences \cite{geeng_egregor_2020, prather_its_2023}. For example, designers who emphasize the competence of a system may unintentionally worsen user experiences in collaborative tasks with the AI \cite{khadpe_conceptual_2020} because of an increase in perceived threat via the co-activation of autonomy and competence mental models. AI governance and policymaking should also consider the public’s mental models of AI risk and regulation \cite{bullock_public_2025}. Without such consideration, they risk endangering people by neglecting desirable safety regulations, and the general rejection of AI systems that would otherwise be beneficial to society such as environmental modeling systems that offer more rapid and effective solutions to climate change.

For organizations, policies allowing human teammates to tune or personalize their AI teammates’ level of autonomy \cite{hauptman_adapt_2023, munyaka_decision_2023} can be informed by our disentanglement of users’ mental models of AI autonomy and sentience. Guidelines for best practice policies should consider mental models. Additionally, calls for corporate policies safeguarding user experiences with companion chatbots (e.g., Replika, ChatGPT) suggest that more responsibility be placed on designers and organizations to align AI systems to users’ mental models because of poor user experiences such as needing to train the chatbot or feeling unsafe during chatbot interactions \cite{namvarpour_uncovering_2024}. Negotiating mental models through mutual theory of mind offers a promising design approach, as evidenced in studies of human-AI cooperation \cite{he_interaction_2023, zhang_towards_2023} suggesting that designs working to align user and system mental models in cooperative tasks increase efficiency, literacy, and ease of use. 

For users, policies that enable personalization based on their own mental models might improve experiences with AI systems \cite{desolda_empowering_2017}, although frequently changing AI features could undermine trust and transparency over time \cite{kahr_good_2025}. Our results can also inform policy standards for prompt engineering in conversational systems \cite{kim_exploring_2025} to account for users’ mental models of autonomy and sentience. Prompts that imply sensing, information processing, emotions, moods, and other affective states are likely to generate perceptions of sentience. Prompts that imply goals, motivations, independent actions, and making choices are likely to generate perceptions of autonomy. Without policies regulating prompts, users may be easily influenced towards a certain perception and its downstream consequences on trust, threat perception, and moral consideration \cite{djeffal_reflexive_2025}. On the other hand, the creation of standardized literacy materials to educate users about mental models and the responses they might receive to their own prompts could assist in generating better outputs \cite{liu_design_2022}.

\subsection{Limitations and Future Directions}

A primary limitation of experimental research on AI perception is the potential disconnect from real-world interaction. However, real-world interaction studies do not necessarily solve all methodological issues, such as the hypotheticality of AI sentience, whether or not sentience perception is influential in human-AI interaction \cite{ladak_public_2025}, and the extent to which studies on platforms such as Prolific reflect the behavior of the general public \cite{pauketat_aims2024_2026}. With a “science fiction science” method \cite{rahwan_science_2025}, we demonstrated that autonomy consistently increased the perception of sentience across experiments, and that this perception explained moral consideration responses, suggesting that sentience plays a role in HCI regardless of its hypotheticality. Further, studies of public opinion on sentient AI have shown that people already react to AI sentience \cite{anthis_perceptions_2025, pauketat_predicting_2022, pauketat_world-making_2025}, and there is a dearth of evidence explaining these reactions. A related limitation is that mental models of non-autonomous and non-sentient machines (e.g., smart TVs, calculators), autonomous but not sentient AIs (e.g., ChatGPT, Roomba), and autonomous and sentient AIs (e.g., Data from the TV series \textit{Star Trek: The Next Generation} \cite{roddenberry_star_1987}, Samantha from the film \textit{Her} \cite{jonze_her_2013}) are easier to imagine than sentient but not autonomous AIs (e.g., organoid intelligences). 

More research is needed to disentangle the effects of autonomy and sentience—in different forms such as hypothetical and actual—from other variables such as anthropomorphism and beliefs about AI souls, personhood, or “magic” \cite{tully_lower_2025}. Relying on a nebulous and shifting concept of “human-likeness” may obscure meaningful explanations caused by capacities of human-likeness (e.g., autonomy, sentience) that are activated, and interacted with, differently in different contexts. For instance, people might react differently to a human-like robot that has experienced harm than to the same human-like robot making a decision requiring trust because different mental models are activated by these contexts (i.e., sentience and autonomy, respectively). Likewise, the prosociality of AI systems is important \cite{ladak_which_2024}, but the levers increasing judgments of AI prosociality are understudied beyond avoiding the uncanny valley with a Goldilocks level of similarity to humans or biological companion dogs \cite{konok_should_2018, yam_reducing_2021}.

The social, moral, and organizational contexts in which AI systems are embedded, their physical forms and capacities, and their degree of interaction with humans will continue to change. Investigating whether mental models of autonomy or sentience interact with any of these, or with mental models of other capacities (e.g., language, intelligence) is needed. For instance, \citet{stein_matter_2020} found that people felt eerie towards a complex AI regardless of their human-like embodiment whereas a simple algorithm was only uncanny when embedded in a human body. Considering reactions to intersecting mental models, multifaceted, and multi-context systems could support more nuanced and detailed scientific theories of human-AI interaction, especially with anthropomorphized AI systems. Finally, we have a unique opportunity to document user reactions to an emerging social category of “digital minds” \cite{anthis_perceptions_2025, kahn_new_2011, severson_behaving_2010, weisman_extraordinary_2022}—AI systems that have or seem to have mental capacities—resulting from technological advances and the fundamental social cognitive perception of computers as social actors with minds.

\section{Conclusions}

Mental models shape human-AI interaction. Our experiments spotlight reactions to autonomy and sentience, two powerful capacities, the perceptions of which have implications for design and policymaking. While studying these reactions may seem speculative, more and more people perceive AI systems like agentic web browsers and emotive companions through their mental models of autonomy and sentience, and heterogeneous, advanced AI systems are becoming embedded in a growing variety of social contexts. Rigorously operationalizing and experimentally studying these mental models provides a foundation for advances in the study of HCI and human-AI interaction.

\begin{acks} 

We would like to thank Jon Bogard, Elisabeth Bradford, Michael Dello-Iacovo, Kurt Gray, Jamie Harris, Ali Ladak, Brad Saad, and Matti Wilks for their thoughtful feedback. 

\end{acks}

\bibliographystyle{ACM-Reference-Format}
\bibliography{ASpaperCHIreferences}

\section*{Online Resources}  The data and materials that support this paper are available on the OSF.
{\small
\begin{itemize}     
    \item \textbf{Pilot studies}     
    \begin{itemize}         
        \item{preregistration} (\url{https://osf.io/rhf6z})
        \item{data and materials} (\url{https://osf.io/r6xyg/})
    \end{itemize}
    \item \textbf {Experiment 1} 
     \begin{itemize}
        \item{preregistration} (\url{https://osf.io/ewxjt})
        \item{data and materials} (\url{https://osf.io/aqexn/})
    \end{itemize}
    \item \textbf{Experiment 2} 
     \begin{itemize}
        \item{preregistration} (\url{https://osf.io/ztmqg})
        \item{data and materials} (\url{https://osf.io/3ua2v/})
     \end{itemize}
    \item \textbf{Experiment 3}
     \begin{itemize}
        \item{preregistration} (\url{https://osf.io/z86gr})
        \item{data and materials} (\url{https://osf.io/wfgk2/})
     \end{itemize}
    \item \textbf {Experiment 4} 
    \begin{itemize}
        \item{preregistration} (\url{https://osf.io/s698b})
        \item{data and materials} (\url{https://osf.io/mquvw/})
    \end{itemize}
    \item \textbf {Supplementary materials, results, and meta-analysis} 
    \begin{itemize}
        \item{\url{https://osf.io/fbt3r/}}     
        \end{itemize} 
        \end{itemize} 
}        
\end{document}